\documentclass[12pt,journal,onecolumn,draftclsnofoot]{IEEEtran}
\newif\ifCLASSOPTIONromanappendices \CLASSOPTIONromanappendicestrue
\usepackage{tikz}
\usepackage[siunitx]{circuitikz}
\usepackage[utf8]{inputenc}
\usepackage{fontenc}
\usepackage{hyperref}
\usepackage{a0size}         
\usepackage{amssymb}
\usepackage{amsmath}
\usepackage{xfrac}
\usepackage{multicol}
\usepackage[english]{babel}    
\usepackage{epsfig} 
\usepackage{bm}
\usepackage{oplotsymbl}
\usepackage{bbm}
\usepackage{amsfonts,color,amsthm,amsmath, lscape,graphics}
\usepackage{url}
\usepackage{comment}
\usepackage{algorithm} 
\usepackage{algpseudocode}

\newtheorem{lemma}{Lemma}
\newtheorem{result}{Result}
\usepackage[font=footnotesize]{caption}
\usepackage[font=footnotesize]{subcaption}
\usepackage{subcaption}
\usepackage{epstopdf}
\usepackage{algorithm} 
\usepackage{algpseudocode}
\usepackage{cite}
\usepackage{subfiles}
\usepackage[colorinlistoftodos]{todonotes}
\definecolor{awesome}{rgb}{1.0, 0.13, 0.32}
\usepackage{fix2col}

\addto\captionsenglish{}

\usepackage{graphicx}  
\theoremstyle{plain}

\usepackage{multirow} 
\usepackage{relsize}  
\usepackage{tabularx}
\usepackage{bm}   
\usepackage{pifont}
\usepackage{etoolbox}

\usetikzlibrary{math}
\tikzmath{\trans = 5;}
\newcommand*\circled[1]{\tikz[baseline=(char.base)]{
            \node[shape=circle,draw,inner sep=2pt] (char) {#1};}}

\makeatletter
\patchcmd{\@maketitle}
  {\addvspace{0.5\baselineskip}\egroup}
  {\addvspace{-2.0\baselineskip}\egroup}
  {}
  {}
\makeatother

\begin{document}
\title{Achievable Rate with Antenna Size Constraint: Shannon meets Chu and Bode}
\author{Volodymyr Shyianov\IEEEauthorrefmark{1}, Mohamed Akrout\IEEEauthorrefmark{1}, Faouzi~Bellili, \IEEEmembership{Member, IEEE}, \\ Amine Mezghani, \IEEEmembership{Member, IEEE}, Robert W. Heath, \IEEEmembership{Fellow, IEEE}
\thanks{\scriptsize{V.~Shyianov, M.~Akrout, F.~Bellili, and A.~Mezghani are with the ECE Department at the University of Manitoba, Winnipeg, MB, Canada (emails:  \{shyianov, akroutm\}@myumanitoba.ca, \{Faouzi.Bellili, Amine Mezghani\}@umanitoba.ca). R.~W.~Heath was with  The University of Texas at Austin and is now at the North Carolina State University (email: rwheathjr@ncsu.edu). This work was supported by the Natural Sciences and Engineering Research Council of Canada (NSERC) and the US National Science Foundation (NSF) Grant No. ECCS-1711702 and CNS-1731658. $^*$Equal contribution. }}}
\maketitle
\begin{abstract}

Using ideas from Chu and Bode/Fano theories, we characterize the maximum achievable rate over the single-input single-output wireless communication channels under a restriction on the antenna size at the receiver. By employing circuit-theoretic multiport models for radio communication systems, we derive the information-theoretic limits of compact antennas. We first describe an equivalent Chu's antenna circuit under the physical realizability conditions of its reflection coefficient. Such a design allows us to subsequently compute the achievable rate for a given receive antenna size thereby providing a physical bound on the system performance that we compare to the standard size-unconstrained Shannon capacity. We also determine the effective signal-to-noise ratio (SNR) which strongly depends on the antenna size and experiences an apparent finite-size performance degradation where only a fraction of Shannon capacity can be achieved. We further determine the optimal signaling bandwidth which shows that impedance matching is essential in both narrowband and broadband scenarios. We also examine the achievable rate in presence of interference showing that the size constraint is immaterial in interference-limited scenarios. Finally, our numerical results of the derived achievable rate as function of the antenna size and the SNR reveal new insights for the physically consistent design of radio systems.
\end{abstract}

\section{Introduction}
\subsection{Background and Motivation}
The analysis and design of communication systems involve a broad spectrum of scientific disciplines ranging from electromagnetic field theory to communication theory to information theory. The physics of radio communication has been captured in the study of antenna and radio-frequency (RF) engineering. Information theory is an abstract mathematical theory guiding the design and implementation of communication systems. Most of the modern communication theory has evolved around the seminal work of Shannon \cite{shannon1948mathematical}, particularly the capacity of the band-limited AWGN channel. The separation between the physical and mathematical abstractions of communication theory has proved convenient since the two are entirely based on a different set of scientific principles. With many new applications driving the demand for higher frequencies, wider bandwidths, and more compact antennas, it is not possible anymore to keep a clean separation without losing essential insights. Merging the well-established fields of information theory and electromagnetic field theory has led to the development of new paradigms such as the \textit{wave theory of information} \cite{franceschetti2017wave}, \textit{electromagnetic information theory} \cite{gruber2008new,migliore2008electromagnetics}, and \textit{circuit theory  for communication} \cite{ivrlavc2010toward}. The variety of studies in wave radiation and propagation systems has shown that circuit and electromagnetic field theories are essential for the  analysis and design of multiple-input and multiple-output (MIMO) communication systems
\cite{ivrlavc2010toward,wallace2004mutual}. Most of these studies, however, are limited to narrowband communication, and further research is still required to ultimately characterize the physical limitations of wireless systems.  

Information theory for MIMO wireless systems is well understood from many perspectives \cite{heath2018foundations}. With the advent of massive MIMO \cite{marzetta2010noncooperative}, the number of base station antennas was allowed to grow large in multi-user scenarios. With cellular handheld and portable communication devices gaining importance in our everyday life, RF and antenna engineers have focused on designing compact antennas \cite{wong2004compact}. A limited number of studies though have taken advantage of the realizability constraints from both physics and circuit theory viewpoints to establish physically consistent models for wireless  systems \cite{ivrlavc2010toward,wallace2004mutual}. In this context, the effect of mutual coupling on MIMO arrays has been analyzed in \cite{wallace2004mutual} by deriving the mutual information\footnote{James Massey mentioned in \cite{massey1998applied} an interesting definition of the channel as the part of the system we are “unwilling or unable to change”. In this sense, the antenna should not be regarded as part of the channel but rather as a physical constraint that is under our control to a certain extent. To avoid any confusion, we will use the term achievable rate instead of capacity throughout the paper.} of MIMO systems where the mutual coupling accounts for the radiated power constraint, receive matching network as well as array scattering parameters. In \cite{taluja2010information}, the data rate of a two broadband antenna system was derived, along with its matching network characteristic, while incorporating the broadband matching limitations. A more rigorous approach has been developed in \cite{ivrlavc2010toward} to cover not only the physical model of mutual antenna coupling \cite{wallace2004mutual} and the transmit/receive impedance matching \cite{taluja2010information}, but also the physics of signal generation and noise modeling. The circuit models developed in \cite{ivrlavc2010toward} for wireless systems incorporate both the intrinsic noise, originating from the receive amplifiers and the extrinsic noise that is received by the antennas. On the same note, it was shown in \cite{mezghani2018information} that the achievable rate with a sufficiently large antenna array under the total radiated power constraint is mainly limited by the fundamental trade-off between the analog beamforming gain and signal bandwidth. More recently, an information-theoretic methodology for analysis and design of broadband antenna systems was proposed in \cite{saab2019capacity1,saab2019capacity}, as opposed to the conventional methodology that relies on frequency-dependent conjugate impedance matching which is infeasible for compact wideband antennas \cite{Fano}. While the work in \cite{ivrlavc2010toward,wallace2004mutual,taluja2010information} combined circuit-based models and the usual information-theoretic models, none of them considered the antenna size as a physical constraint in their respective designs.

For electrically small antennas, the authors of \cite{gustafsson2007spectral} analyzed the fundamental limitations from both antenna theory and broadband matching perspectives by deriving lower bounds on the spectral efficiency of a compact MIMO antenna array inserted inside a sphere. In this work, we also investigate the performance characterization of electrically small antennas, albeit from an information-theoretic perspective. 
The use of Shannon theory or narrow-band theory  is particularly limiting for the IoT devices since the compact antenna sizes are a fraction of the carrier wavelength, i.e., $a \ll \lambda_c$. This can correspond to an unfortunate design scenario where the size is not dictated by the physical limitations, but rather by antennas-business-related developments, e.g., antennas in mobile phones must have compact sizes due to the aesthetic and form factor requirements. Despite the electrically small size and the matching losses, the
performance is still adequate at  ultrahigh-frequency (UHF) bands which are used in 5G networks.

\subsection{Contributions}
We derive the achievable rate on the single-input single-output (SISO) wireless communication channel with a restriction on the size of the antenna at the receiver only. Using a physically consistent circuit model of radio communication, we find the maximum mutual information per unit time between the input and output signals of the system under the antenna size constraint. This restriction is incorporated into the circuit model by the use of Chu theory \cite{chu1948physical}. The mutual information is optimized with respect to the matching network (MN) between the antenna and the low-noise amplifier (LNA). Broadband matching theory \cite{Fano,bode1930method} is further leveraged to obtain a physically realizable MN. For a given size of the receiver antenna structure, the mutual information is computed and compared to the standard size-unconstrained Shannon capacity. It is found that the received SNR is a strong function of the antenna size and that finite-size performance degradation is most apparent in the low-SNR regime where only a small fraction of Shannon capacity can be achieved. We extend the SISO results to find the mutual information in presence of interference. We show that the finite-size performance degradation vanishes in interference-limited scenarios (when interference dominates amplifier noise). Moreover, we determine the optimal signaling bandwidth thereby showing that impedance matching can offer a substantial performance improvements. Based on the mutual information optimization methodology, we demonstrate that the optimal matching network has a non-flat frequency response in the pass-band, unlike the more conventional flat frequency response assumed in the analysis of small antennas \cite{gustafsson2007spectral}. While standard filter design methodologies \cite{pozar2000microwave} might be insufficient to obtain a good approximation to the optimal matching network obtained in this work, we leave alternative synthesis techniques for future investigation.
\subsection{Paper Organization and Notation}
We structure the rest of this paper as follows. In Section~\ref{section_2}, we introduce the communication model, along with its equivalent channel circuit based on the antenna limitations and broadband matching theory by considering the physical constraints established by using Chu and Bode/Fano theories. In Section~\ref{section_4}, we derive the achievable rate of the resulting channel model given in the form of a parametric equation involving the antenna-size. We then extend the computation of the achievable rate to the homogeneous interference scenario in Section~\ref{sec:interference} by approximating the interference power density. Finally, in Section~\ref{section_6}, we solve the parametric equations numerically, from which we draw out concluding remarks in the presence or absence of interference.
\newline 
\indent The following notation is used in this paper. 
   Given any complex number,  $\Re\{\cdot\}$, $\Im\{\cdot\}$, and $\{\cdot\}^*$ return its real part, imaginary part, and complex conjugate, respectively.  The statistical expectation and variance are denoted as $\mathbb{E}[\cdot]$ and $\textrm{Var}[\cdot]$, respectively. We also denote $j$ as the imaginary unit (i.e., $j^{2}=-1$). Throughout the paper, $c$ denotes the speed of light in vacuum (i.e., $c \approx 3\times10^8\,[\textrm{m}/\textrm{s}]$), $T$ is the temperature in Kelvin, $\lambda$ is the wavelength, $k$ is the wave number, and $k_\text{b} = 1.38 \times 10^{-23}\, [\textrm{m}^{2}\, \textrm{kg} \,\textrm{s}^{-2}\, \textrm{K}^{-1}]$ is the Boltzmann constant. $\mu = 1.25\times 10^{-6} \,[\textrm{m}\, \textrm{kg} \,\textrm{s}^{-2}\, \textrm{A}^{-2}]$ and $\epsilon = 8.85\times 10^{-12} \,[\textrm{m}^{-3}\, \textrm{kg}^{-1} \,\textrm{s}^{4}\, \textrm{A}^{2}]$ are the permeability and permittivity of vacuum.

\section{System Model}\label{section_2}
When choosing an appropriate tool for analyzing communication systems, it is essential to consider the interface between information and antenna theories. Applying electromagnetic (EM) field theory directly to communication problems is a difficult endeavour~\cite{ivrlavc2010toward}. A circuit-based model for analyzing communication
systems, which is consistent with the governing laws of physics, is relatively much simpler.
\subsection{Circuit model of a communication system}
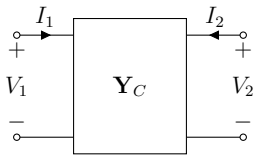
\begin{figure}[h!]
\centering
\begin{circuitikz}[american voltages, american currents, scale=0.75, every node/.style={transform shape}]
\draw (0,0)
node[draw,minimum width=2cm,minimum height=2.4cm] (load) {$\mathbf{Y}_C$}
  ($(load.west)!0.75!(load.north west)$) coordinate (la)
  ($(load.west)!0.75!(load.south west)$) coordinate (lb)
  ($(load.east)!0.75!(load.north east)$) coordinate (lc)
  ($(load.east)!0.75!(load.south east)$) coordinate (ld)
  
  ($(lb) + (-1,0)$) to[short,o-] (lb)
  ($(la) + (-1,0)$) to[short,i>^=$I_1$,o-] (la)
  ($(ld) + (1,0)$) to[short,o-] (ld)
  ($(lc) + (1,0)$) to[short,i>_=$I_2$,o-] (lc)
  
  ($(la) + (-1,0)$) to [open,v=$V_1$] ($(lb) + (-1,0)$)
  ($(lc) + (1,0)$) to [open,v^=$V_2$] ($(ld) + (1,0)$)
;\end{circuitikz}
\caption{A communication channel modeled as a two-port network with the channel input port ($V_1$, $I_1$)  and the channel output port ($V_2$, $I_2$).}
\label{fig:two-port}
\end{figure}

 A communication channel can be viewed as a black-box establishing the relationship between the port voltages, $(V_1(f),V_2(f))$, and port currents, $(I_1(f),I_2(f))$, through a symmetric admittance matrix $\mathbf{Y}_C$ as depicted in Fig.~\ref{fig:two-port}.
  When a generator is connected to the port of the transmitting antenna, the current flow on the antenna surface generates an EM field in the space outside the antenna structure (i.e., the generator together with the antenna and the transmission line connecting them). Similarly, the reception of the EM signal impinging on the receive antenna is manifested by a voltage induced on the antenna port and a current flow on the antenna surface.
 From circuit theory, establishing the relationship between port variables is all that is necessary to consistently model the single-input single-output (SISO) communication channel as an \textit{electrical two-port network}.

In information theory, the information-carrying transmitted signals are random processes. From this standpoint, it is insightful to study the mutual information of any physical volume (in space) used for receiving (or generating) these signals. By virtue of their simplicity, circuit-theoretic tools appear to be well suited to such information-theoretic analysis as will be demonstrated later on.
\subsection{Achievable rate of a wireless communication channel}
The achievable rate  of the continuous-time additive Gaussian noise channel with a certain transmit power spectral density (PSD) $P_{\rm t}(f)$, a channel frequency response $H(f)$, and a noise PSD $N(f)$, is given by \cite{gallager1968information}:
\begin{equation}\label{AWGN_Capacity}
    C_{[\textrm{bits/s}]} = \int_{0}^{\infty}{\log_2\left(1+\frac{P_{\rm t}(f)|H(f)|^2}{N(f)}\right)\textrm{d}f}.
\end{equation}
Taking $|H(f)|^2 = 1$, $N(f) = N_0$ and uniform power allocation $P_t(f) = P/\textrm{BW}$ across a certain bandwidth $\textrm{BW}$, one recovers the well-known capacity of the AWGN channel from (\ref{AWGN_Capacity}). In wireless communication, it is more common to include the path-loss in the form of the Friis' transmission equation
\cite{heath2018foundations}. If $d$ is the distance between the transmitter and reciever and $G_\text{t}$ and $G_\text{r}$ are recieve and transmit antenna gains respectively, the Friis' transmission equation takes the form 
\begin{equation}\label{channel_squared_magnitude}
    |H(f)|^2 = G_\text{t}\,G_\text{r} \left(\frac{c}{4{\pi}fd}\right)^2,
\end{equation}
 By the simple application of (\ref{channel_squared_magnitude}), the signal part of the received power spectrum can then be computed from
\begin{equation}
    P_r(f) = P_{\rm t}(f)\,G_\text{t}\,G_\text{r}\left(\frac{c}{4{\pi}fd}\right)^2\, \bigg[\frac{\textrm{W}}{\textrm{Hz}}\bigg].
\end{equation}
\subsection{The antenna size constraint}\label{sec: Chu stuff}
Suppose that the receive antenna structure is embedded inside a geometrical spherical volume of radius $a$. Physical intuition suggests that the data rate should depend on the antenna size, it might be convincing to argue that both the signal and noise powers would be affected in the same way as the antenna size is decreased thereby leaving the SNR, $\textrm{SNR}(f) = \frac{P_r(f)}{N(f)}$, unchanged.
Indeed, while decreasing the antenna size affects the received signal and background radiation noise in the same way, the intrinsic amplifier noise of a realistic receiver would, in turn, make the $\textrm{SNR}(f)$ or equivalently the achievable rate go to zero. As will be shown in the subsequent sections, the achievable rate in (\ref{AWGN_Capacity}) holds approximately only when $\textrm{BW}\ll f_{\textrm{c}}$ where $\textrm{BW}$ is the absolute bandwidth of the transmitted signal and $f_{\textrm{c}}$ is the carrier frequency (in a narrowband system). Although this argument supports the intuition that there should exist a definite relationship between the achievable rate and the receive antenna size, it does not offer a way of obtaining such a relationship. To answer the question of how to find the maximum achievable performance of any antenna structure with a given size, it necessary to resort to EM field theory. Fortunately, as per Chu's seminal work \cite{chu1948physical}, any antenna structure that can be embedded inside a spherical volume, of a given radius $a$, can be represented by an equivalent circuit model corresponding to the superposition of $\textrm{TM}_\text{n}$ radiation modes.

\begin{figure}[b]
\centering
\begin{circuitikz}[scale=0.75, every node/.style={transform shape}]
\draw (0,0) node[anchor=east]{}
 to[short, o-*] (3,0)
 (3,2) to[L, label=\mbox{\small{$L=\frac{a \,R_2}{c}$}}, *-*] (3,0)
 (3,0) -- (5,0)
 (5,2) to[/tikz/circuitikz/bipoles/length=30pt, R, l=\mbox{\small{$R_2$}}, -] (5,0)
 (3,2) -- (5,2)
 (0,2) node[anchor=east]{}
  to[C, i>_=\mbox{\small{$I_1(f)$}}, label=\mbox{\small{$C=\frac{a}{cR_2}$}}, o-*] (3,2);
  
  \draw[-latex] (0,0.5) -- node[above=0.05mm] {$Z_1(f)$} (1, 0.5);
\end{circuitikz}
\caption{Equivalent circuit of the TM$_1$ mode.}
\label{fig:tm1}
\end{figure}
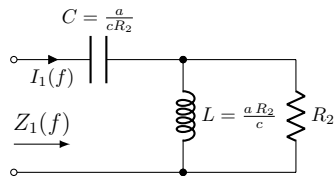

In mobile communication, where omnidirectional antenna patterns are most preferable, it is enough to consider the first mode of radiation only, i.e., $n=1$. In this case, the equivalent  circuit for the TM$_1$ wave corresponds to the so-called ``\textit{electric Chu's antenna}'' and is illustrated in Fig.~\ref{fig:tm1}  which represents its equivalent circuit model at the receiver with
\begin{subequations}\label{appendix-eq:V-I-TM1}
    \begin{align}
    V_{1}&~=~j\,\sqrt[4]{\frac{\mu}{\epsilon}} \,\frac{\sqrt{R_2}A_{1}}{k} \,\sqrt{\frac{8 \pi}{3}} \, \frac{\partial}{\partial\rho} \big(\rho\, h_1(\rho)\big) \,[\textrm{V}],\\
I_{1}&~=~\sqrt[4]{\frac{\mu}{\epsilon}} \,\frac{A_{1}}{\sqrt{R_2}\,k}\, \sqrt{\frac{8 \pi}{3}} \,\rho \,h_{1}(\rho)\,[\textrm{A}],\\
Z_1 &~=~ jR_2\,\Bigg(\frac{1}{\rho} + \frac{1}{h_1(\rho)}\, \frac{\partial}{\partial\rho} h_1(\rho)\Bigg)\, [\Omega].
    \end{align}
\end{subequations}
where $\rho = ka$.
Using the recursion for the spherical Bessel functions, one can write the input impedance $Z_1$ as:
\begin{equation}\label{eq:Z1-chu}
\begin{aligned}
    Z_1 ~=~\underbrace{\frac{R_2}{j\,2\pi f\,\frac{a}{c}}}_{Z_C} + \underbrace{\frac{R_2}{\frac{1}{j\,2 \pi f\,\frac{a}{c}}+1}}_{Z_{\rm L} \,//\, Z_R}\, [\Omega],
    \end{aligned}
\end{equation}
which represents the impedance of the  circuit illustrated in Fig.~\ref{fig:tm1}. The driving point impedance (\ref{eq:Z1-chu}) of the electric Chu's antenna is a special case (when $n=1$) of their general expressions for any radiation mode $\textrm{TM}_{\textrm{n}}$ as discussed in \cite{chu1948physical}. It should be noted that the antenna structure which excites only the $\textrm{TM}_\text{1}$ mode outside the sphere has the broadest bandwidth\footnote{A better bandwidth could be achieved (i.e, improved Chu limit) if we combine TM$_1$ and TE$_1$, known as magneto-electric antenna \cite{hansen2011small}, for simplicity consideration is given to the electric antenna only.} of all antennas with a linearly polarized omnidirectional pattern~\cite{chu1948physical}.
The fact that the electric Chu's antenna exciting the TE$_1$ mode only (with maximum directivity of $\frac{3}{2}$ \cite[chapter 6]{harrington1961pp}) has the broadest bandwidth compared to all the Chu's antennas operating by exciting the higher-order modes allows us to benefit from its simplicity while being consistent with the physical constraints on the antenna size.
From this perspective, we will rely in this paper on the circuit-theoretic model of the Chu's electric antenna depicted in Fig.~\ref{fig:tm1} to obtain the maximum achievable rate for any radio receiver of a fixed size.
\subsection{Physically realizable impedance matching}
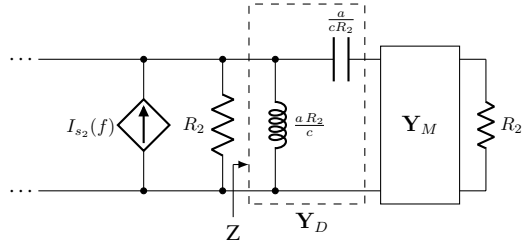
\begin{figure}[h!]
\centering
\begin{circuitikz}[scale=0.7,american voltages, american currents, every node/.style={transform shape}]
\draw (2.5,2.5) to[short] (4.5,2.5);
\draw (2.5,0) to[short] (4.5,0);
\draw (4.5,0) to[cI,*-*] (4.5,2.5)
(4.5,2.5) to[short] (5,2.5)
(4.5,0) to[short] (5,0);
\draw (6.5,2.5) to [short] (5,2.5)
(6.5,0) to [short] (5,0)
(6.5,2.5) to[short] (7.5,2.5)
(6,0) to[R, label=\mbox {\small{$R_2$}}, *-*] (6,2.5)
(6.5,0) to[short] (7.5,0)
(7,2.5) to[L, label=\mbox{\small{$\frac{a \,R_2}{c}$}}, *-*] (7,0)
(7.5,0) to[short] (9,0)
(7.5,2.5) to[C, label=\mbox{\small{$\frac{a}{cR_2}$}}] (9,2.5);
\draw (9,-0.25) rectangle (10.5,2.75) node[pos=.5] {$\mathbf{Y}_M$};
\draw (10.5,2.5) to[short] (11,2.5);
\draw (10.5,0) to[short] (11,0)
(11,2.5)  to[/tikz/circuitikz/bipoles/length=30pt,R, l^=\mbox{\small{$R_{2}$}}, -] (11,0);
\draw[dashed] (6.5,-0.25) rectangle +(2.2,3.8);
\node[] at (3.5,1.2) {\small{$I_{s_2}(f)$}};
\node[] at (7.7,-0.6) {$\mathbf{Y}_D$};
\draw[-latex] (6.2,0.5) -- node[above=0.05mm] {} (6.5, 0.5);
\draw (6.2,0.5) to[short] (6.2,-0.5);
\node[] at (6.2,-0.8) {Z};
\node[] at (2.2,2.5) {\dots};
\node[] at (2.2,0) {\dots};
\end{circuitikz}
\caption{Matching circuit with Chu antenna.}
\label{fig:Fano_matching} 
\end{figure}

It is known that the bandwidth of an antenna can be improved by incorporating an impedance matching network between the antenna and an amplifier. The problem of matching an arbitrary load impedance $Z_{\rm L}$ to a purely resistive source was addressed in \cite{Fano}. We review the general principles of broadband matching theory and specialize it to the antenna model at hand in Appendix \ref{Fano_appendix}. By closely inspecting the circuits in Fig.~\ref{fig:two-port_channel} and \ref{fig:channel_model}, the matching problem can be stated with reference to Fig.~\ref{fig:Fano_matching} as follows. Find the conditions of physical realizability of the input impedance, $Z(f)$, or equivalently the input reflection coefficient, $\Gamma(f)$, whose magnitude    
\begin{equation}
|\Gamma(f)|^2 ~=~ 1 - \frac{P_L(f)}{P_{\text{max}}(f)}
\end{equation}
 corresponds to the fraction of power being rejected by the LNA.
Using Darlington representation of Chu-equivalent circuit as we show in Appendix \ref{Fano_appendix}, the conditions of physical realizability of the reflection coefficient take the form of the following two integral constraints:
\begin{equation}\label{constraint_1}
\frac{1}{2\pi^2} \int_{0}^{\infty} f^{-2} \ln\Bigg( \frac{1}{|\Gamma(f)|^2}\Bigg) \,\mathrm{d}f~=~\left(\frac{2 a}{c}-2 \gamma^{-1}\right),
\end{equation}
and
\begin{equation}\label{constraint_2}
\frac{1}{8\pi^4} \int_{0}^{\infty} f^{-4} \ln\Bigg( \frac{1}{|\Gamma(f)|^2}\Bigg) \,\mathrm{d}f~=~\left(\frac{4 a^{3}}{3 c^{3}}+\frac{2}{3} \gamma^{-3}\right),
\end{equation}
where $\gamma$ is the positive real-valued zero of the reflection coefficient.\\
We have covered the background on the electric Chu's antenna and its respective optimal transmission coefficient, which will lay the ground for the maximization of the achievable rate of the SISO channel in Section \ref{subsec:optimization-methodology} once the SISO channel is fully characterized. This will be
covered in the next section.
\subsection{A circuit-theoretic SISO communication model}
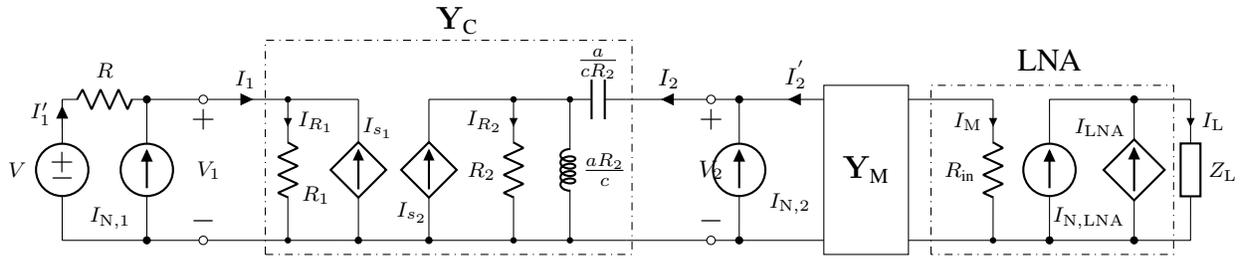
\begin{figure}[h!]
\centering
\begin{circuitikz}[scale=0.75,american voltages, american currents]
\draw (0-\trans,2) to[short, i>^=\scriptsize{$I_1'$}] (0-\trans,2.5);
\draw (0-\trans,2.5) to[/tikz/circuitikz/bipoles/length=33pt, V, l_=\scriptsize{$V$}] (0-\trans,0)
(0-\trans,0) to[short,-o] (2.5-\trans,0)
(0-\trans,2.5) to [/tikz/circuitikz/bipoles/length=25pt,R,l=\scriptsize{$R$}, -*] (1.5-\trans,2.5)
(1.5-\trans,0) to[/tikz/circuitikz/bipoles/length=33pt, I, l_=\mbox{},*-*] (1.5-\trans,2.5)
(1.5-\trans,2.5) to[short, -o] (2.5-\trans,2.5)
(2.5-\trans,2.5) to [short,i>^=\scriptsize{$I_1$}, -] (4-\trans,2.5)
(2.5-\trans,2.5) to [open,v=\small{}] (2.5-\trans,0)
(2.5-\trans,0) to[short] (4-\trans,0);
\draw (4-\trans,2.5) to[/tikz/circuitikz/bipoles/length=25pt,R,i>_=\mbox{},*-*] (4-\trans,0);
\draw (5.25-\trans,0) to[/tikz/circuitikz/bipoles/length=30pt,cI, label=\mbox{}, *-] (5.25-\trans,2.5);
\draw (4-\trans,0) to[short] (5-\trans,0);
\draw (4-\trans,2.5) to[short] (5.25-\trans,2.5);
\draw (5-\trans,0) to[short] (6.5-\trans,0);
\draw (6.5-\trans,0) to[/tikz/circuitikz/bipoles/length=30pt,cI, label=\mbox{}, *-] (6.5-\trans,2.5);
\draw (8-\trans,2.5)  to[/tikz/circuitikz/bipoles/length=25pt,R, l_=\mbox{\scriptsize{$R_2$}}, i>_=\mbox{\scriptsize{$I_{R_2}$}}, *-*] (8-\trans,0);
\draw (6.5-\trans,0) to[short] (8-\trans,0);
\draw (6.5-\trans,2.5) to[short] (8-\trans,2.5);
\draw (9-\trans,0)  to[L, /tikz/circuitikz/bipoles/length=25pt, l_=\mbox{\small{$\frac{a R_2}{c}$}}, *-*] (9-\trans,2.5);
\draw (8-\trans,2.5) to[short] (9-\trans,2.5);
\draw (8-\trans,0) to[short] (9-\trans,0);
\draw (9-\trans,2.5) to[C, /tikz/circuitikz/bipoles/length=25pt, label=\mbox{\small{$\frac{a}{c R_2}$}}] (10-\trans,2.5);
\draw (9-\trans,0) to[short] (10-\trans,0);
\draw (6.5,2.5) to [short,i>_=\footnotesize{$I_2$}, o-] (5,2.5)
(6.5,0) to [short, o-] (5,0)
(6.5,2.5) to [open,v_=\footnotesize{$V_2$}] (6.5,0)
(6.5,0) to[short] (7.5,0)
(6.5,2.5) to[short] (7.5,2.5)
(7,0) to[/tikz/circuitikz/bipoles/length=33pt, I, l_=\mbox{},*-*] (7,2.5)
(7.5,0) to[short] (8.5,0)
(8.5,2.5) to[short,i>_=\footnotesize{$I^{'}_2$}] (7.5,2.5);
\draw (8.5,-0.25) rectangle (10,2.75) node[pos=.5] {$\mathbf{Y}_{\textrm{M}}$};
\draw (10,2.5) to[short] (11.5,2.5);
\draw (10,0) to[short] (11.5,0)
(11.5,2.5)  to[/tikz/circuitikz/bipoles/length=25pt,R,i>_={\scriptsize $I_{\rm{M}}$}, l_=\mbox{\scriptsize{$R_{\textrm{in}}$}}, -*] (11.5,0)
(11.5,0) to[short] (12.5,0)
(12.5,0) to[/tikz/circuitikz/bipoles/length=33pt, I, l_=\mbox{},*-] (12.5,2.5)
(12.5,0) -- (14,0)
(14,0) to[/tikz/circuitikz/bipoles/length=33pt, cI, label=\mbox{},*-*] (14,2.5)
(12.5,2.5) -- (14,2.5)
(14,2.5) to[short] (15,2.5)
(15,2.5) to[european resistor, /tikz/circuitikz/bipoles/length=25pt, i>^=\scriptsize{$I_{\rm{L}}$}, label=\mbox{\scriptsize{$Z_{\rm{L}}$}},-] (15,0)
(14,0) -- (15,0);
\node[] at (13.2,0.4) {{\scriptsize $I_{\rm{N}, \rm{LNA}}$}};
\node[] at (13.4,2) {{\scriptsize $I_{\rm{LNA}}$}};
\node[] at (7.9,0.65) {{\scriptsize $I_{\rm{N},2}$}};
\node[] at (2.6-\trans,1.25) {\scriptsize{$V_1$}};
\node[] at (0.85-\trans,0.4) {\scriptsize{$I_{\rm{N},1}$}};
\node[] at (5.6-\trans,2) {\scriptsize{$I_{s_1}$}};
\node[] at (4.5-\trans,2.1) {\scriptsize{$I_{R_1}$}};
\node[] at (6.2-\trans,0.5) {\scriptsize{$I_{s_2}$}};
\node[] at (4.5-\trans,0.8) {\scriptsize{$R_1$}};
\node[] at (7-\trans,3.9) {$\mathbf{Y}_{\textrm{C}}$};
\node[] at (17.5-\trans,3.2) {$\textrm{LNA}$};
\draw[dash dot] (10.4,-0.25) rectangle +(4.3,3);
\draw[dash dot] (3.6-\trans,-0.25) rectangle +(11.5-\trans,3.8);
\end{circuitikz}

\caption{SISO communication model from left to right: the signal generator $V(f)$ and its resistance $R$, the extrinsic noise of the transmit antenna $I_{\textrm{N,1}}(f)$, the transmit/receive antenna model $\mathbf{Y}_C$, the extrinsic noise of the receive antenna $I_{\textrm{N,2}}(f)$, the matching network $\mathbf{Y}_{\textrm{M}}$, the receive amplifier model in the dashed box, the load impedance $Z_{\rm L}$ whose current, $I_{\rm{L}}(f)$ constitutes the output signal of the communication system.}
\label{fig:two-port_channel}
\end{figure}

 A circuit-theoretic model of SISO communication systems that includes receive impedance matching, antenna channel and LNA is depicted as in Fig.~\ref{fig:two-port_channel}.
 
 There, the signal generator is represented by the voltage generator $V(f)$ with its internal resistance $R$ in series. The current sources $I_{\textrm{N,1}}(f)$ and $I_{\textrm{N,2}}(f)$ surrounding the transmit/receive antenna model, $\mathbf{Y}_C$, account for the extrinsic noise of the transmit antenna and the receive antenna, respectively. In this regard, the conjugate pairs \big($V_1(f)$, $I_1(f)$\big) and \big($V_2(f)$, $I_2(f)$\big) can be interpreted as the voltage and current of the transmit antenna and the receive antenna, respectively. To deliver the maximum power to the load $Z_{\rm L}$, the matching network $\mathbf{Y}_{\textrm{M}}$ matches the input impedance of the receive antenna to the receive amplifier. The equivalent circuit model of the latter framed in a dashed box represents a current amplifier with an input resistance $R_{\text{in}}$ that accounts
for the fact that the amplifier draws an input current from the matching network, a current-controlled current source, $I_{\textrm{LNA}}(f)$, having a gain factor $\beta$ in parallel to an independent current source $I_{\textrm{N,LNA}}(f)$ modeling the intrinsic noise of the amplifier \cite[chapter 1]{sedra1998microelectronic}.

\noindent The non-ideal generator voltage phasor, $V(f)$, (i.e., with the internal resistance $R$) is a frequency-domain representation of the real pass-band signal, $v(t)$, to be transmitted over the channel. As $v(t)$ is a Gaussian random signal (i.e, not energy limited), it is convenient to use a Fourier transform truncated to the interval of time $T_0$,
\begin{equation}\label{trancated_Fourier}
    V_{T_0}(f) = \int_{-\frac{T_0}{2}}^{+\frac{T_0}{2}}{v(t)\,e^{-2{\pi}jft}}\,\text{d}t\, \bigg[\frac{\textrm{V}}{\textrm{Hz}}\bigg],
\end{equation}
 With the definition in (\ref{trancated_Fourier}), the transmit PSD takes the following form:
\begin{equation}
    P_{\rm t}(f) = \lim_{T_0\to \infty}\frac{1}{T_0}\frac{\mathbb{E}[|V_{T_0}(f)|^2]}{4R}\, \bigg[\frac{\textrm{W}}{\textrm{Hz}}\bigg].
\end{equation}
Notice that the transmit PSD represents the power that would be radiated by the antenna perfectly matched to the source.
\subsubsection{The noisy communication channel model}
The generator terminals are connected to the transmitting antenna through an input port with the current-voltage pair $(I_1(f),V_1(f))$. The channel between the transmit and receive antennas is represented by the frequency-domain admittance matrix $\mathbf{Y}_C$. The receive antenna terminals are connected to the outside world through the output port with the current-voltage pair $(I_2(f),V_2(f))$. The representation constitutes a noiseless two-port network
\begin{equation}
    \Bigg[\begin{array}{l}
I_{1}(f) \\
I_{2}(f)
\end{array}\Bigg]~=~\boldsymbol{Y}_{C}\Bigg[\begin{array}{l}
V_{1}(f) \\
V_{2}(f)
\end{array}\Bigg],
\end{equation}
where the background noise, \big($I_{\textrm{N},1}(f),I_{\textrm{N},2}(f)$\big), is injected at both the input and output ports and its second-order moments are determined from \cite{nyquist} once again using truncated Fourier transform: 
\begin{equation}
    \lim_{T_0\to\infty}\frac{1}{T_0}\,\mathbb{E}[|I_{\rm{N},\mathit{k}}(f)|^2] = 4\,k_\text{b}\,T\,\Re{\{(\mathbf{Y}_C)_{k,k}\}}\, \bigg[\frac{\textrm{A}^{2}}{\textrm{Hz}}\bigg],~~~ k=1,2.
\end{equation}
By applying Kirchhoff's current law (KCL) in Fig.~\ref{fig:two-port_channel}, we obtain an affine noisy two-port communication channel
\begin{equation}
    \Bigg[\begin{array}{l}
I_{1}'(f) \\
I_{2}'(f)
\end{array}\Bigg]~=~\boldsymbol{Y}_{C}\Bigg[\begin{array}{l}
V_{1}(f) \\
V_{2}(f)
\end{array}\Bigg] - \Bigg[\begin{array}{l}
I_{\rm{N},1}(f) \\
I_{\rm{N},2}(f)
\end{array}\Bigg].
\end{equation}
The output port of the channel is further connected to the matching network represented by the lossless two-port network with admittance matrix $\mathbf{Y}_{\textrm{M}}$. The purpose of the matching network is to assure that the maximum amount of mutual information collected by the antenna gets transferred into the LNA at all frequencies that are present in the signal $V(f)$.
\subsubsection{The receive LNA model} The LNA is modeled at the receiver side as a noisy frequency flat device with gain $\beta$,
\begin{equation}
    I_{\rm{LNA}}(f) = \beta \,I_{\rm{M}}(f)\, [\textrm{A}].
\end{equation}
With $R_{\textrm{in}}$ and $N_\textrm{f}$ being the input impedance and the noise factor of the amplifier, respectively. The second-order statistics of the noise current, $I_{\textrm{N,LNA}}(f)$, generated inside the LNA are determined using the truncated Fourier transform 
\begin{equation}\label{eqn:LNA}
    \lim_{T_0\to\infty}\frac{1}{T_0} \,\mathbb{E}[|I_{\rm{N},\rm{LNA}}(f)|^2] = \frac{\beta^2\,k_\text{b}\,T}{R_{\textrm{in}}}\,(N_\text{f} - 1) \, \bigg[\frac{\textrm{A}^{2}}{\textrm{Hz}}\bigg].
\end{equation}
 The definition for the noise factor $N_\text{f}$ in (\ref{eqn:LNA}) is better understood by considering an analogy with a narrowband amplifier matched at both ports. In that narrowband case \cite{pozar2000microwave}, the noise factor corresponds to the noise figure of the receiver under perfect matching. In the broadband case of this work, this ideal noise figure is not attainable under  electrically small antennas and the noise factor $N_\text{f}$ is merely a constant specifying the quality of the LNA and should not be interepreted as noise figure. It is also possible to refine the LNA model by considering two correlated noise sources  at the input port of the LNA \cite{ivrlavc2010toward}. A straightforward extension to a more practical frequency dependent noise factor $N_{\rm f}$ is also possible but just avoided for the sake of simplicity.
\subsection{Our proposed achievable rate optimization methodology}\label{subsec:optimization-methodology}
To find the maximum achievable rate, it is necessary to optimize the mutual information over the parts of the communication system described above that are at the disposal of the system designer. Once the best possible physically realizable design is identified, the resulting mutual information can be interpreted as a supremum of all achievable rates. By revisiting Fig.~\ref{fig:two-port_channel}, it is seen that the transmit waveform, $v(t)$, and the reciprocal lossless matching network, $\mathbf{Y}_{\textrm{M}}$, are under the full control of the system designer. The channel admittance matrix, $\mathbf{Y}_C$, can also be partially designed by obtaining the optimal transmitter and receiver antenna structures. The maximum mutual information is thus given by:
\begin{equation}\label{capacity_1}
    C_{[\textrm{bits}/\textrm{s}]} = \max_{\mathbf{Y}_M,\,\mathbf{Y}_C,\,\mathbb{P}_v} I(v(t);i_{\rm{L}}(t)),
\end{equation}
where $I(v(t);i_{\rm{L}}(t))$ is the mutual information per unit time between the two random processes representing the input and output signals of the communication system\cite{gel},\cite{gallager1968information}. Moreover, $\mathbb{P}_v$ is the probability measure on the space of possible generator voltages, $v(t)$, which for any finite set of time instants $\{t_1,t_2,\ldots,t_n\}$ specifies the joint cumulative distribution function:
\begin{equation}
\mathbb{P}_v[v(t_1)\leq v_1,v(t_2)\leq v_2,\ldots v(t_n)\leq v_n]~~ \forall(v_1,v_2,\ldots,v_n)\in \mathbb{R}^n.
\end{equation}
 In designing the probability law of the generator, we suppose that the expected per-frequency power constraint, $P_{\rm t}(f)$, satisfies:
 \begin{equation}\label{transmit_PSD_constraint}
     P_{\rm t}(f) \leq E_{\textrm{max}},
 \end{equation}
 where $E_{\textrm{max}}$ is the maximum spectral power that the generator is able to supply and is imposed due to regulatory restrictions or hardware constraints.
 
As explained in section \ref{sec: Chu stuff}, the radiation pattern of any antenna structure embedded inside a spherical volume of radius $a$ can be represented by a series of spherical wave functions.  Each mode of radiation can then be equivalently characterized by its current-voltage relationship. The equivalent circuit of the antenna which has only the lowest TM$_1$ mode as the EM field outside the volume is depicted in Fig.~\ref{fig:tm1}, where $c$ is the speed of light in vacuum. The antenna gain for this lowest mode in the equatorial plane is $3/2$ \cite[chapter 6]{harrington1961pp}.

\begin{figure}[h!]
\centering
\begin{circuitikz}[american voltages, american currents, scale=0.7, every node/.style={transform shape}]

\draw (0,0) to[short,-o] (1,0)
(1,0) to[short,o-*] (2.5,0)
(2.5,2) to[/tikz/circuitikz/bipoles/length=30pt,R, i>_=\mbox{\small{$I_{R_1}(f)$}}, l=\mbox{\small{$R_1$}}, *-*] (2.5,0)
(0,2) to [short,i>^=$I_1(f)$, -o] (1,2)
(1,2) to [short,-*] (2.5,2)
(0, 2) to [open,v=$V_1(f)$] (0,0)

(2,0) to[short, -*] (6.5,0)
(4,0) to[cI, label=\mbox{}, *-] (4,2)
(6.5,0) to[cI, label=\mbox{}] (6.5,2)
(2,2) to[short,-] (4,2)
(6.5, 2) to[short,-*] (8,2)
(6.5, 0) to[short,-*] (8,0)
(8,2)  to[/tikz/circuitikz/bipoles/length=30pt,R, l_=\mbox{\small{$R_2$}}, i>_=\mbox{\small{$I_{R_2}(f)$}}, *-*] (8,0)
(8, 2) to[short,-*] (9,2)
(8, 0) to[short,-*] (9,0)
(9,0)  to[L, l_=\mbox{\small{$\frac{a R_2}{c}$}}, *-*] (9,2)

(9,0) to[short,-o] (11,0)
(9,2) to[C, label=\mbox{\small{$\frac{a}{c R_2}$}}, -o] (11,2)
(12.5, 2) to [open,v=$V_2(f)$] (12.5,0)

(12,2) to [short,i>_=$I_2(f)$] (11,2)
(12,0) -- (11,0);
\node[] at (4.7,1.5) {\small{$I_{s_1}(f)$}};
\node[] at (5.8,0.5) {\small{$I_{s_2}(f)$}};
\draw[dashed] (1.25,-0.25) rectangle +(9.5,3.25)

;\end{circuitikz}
\caption{Channel represented as a two-port network.}
\label{fig:channel_model}
\end{figure}
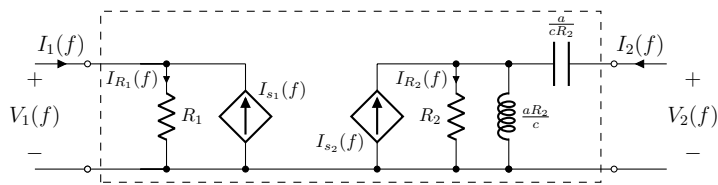

 By inspecting Fig.~\ref{fig:tm1}, any antenna structure of finite size will necessarily have a reactive component present (associated with non-propagating electromagnetic near-fields) in its input impedance. To seek a maximum radiation efficiency, i.e. a purely resistive input impedance, we take the limit as the size $a \rightarrow \infty$, which renders the capacitance a short circuit and the inductance an open circuit.\\\indent
In our present investigation, we will focus solely on the size limitation at the receiver side by restricting the volume embodying the transmit antenna to be of infinite size. This leads to the channel model in Fig.~\ref{fig:channel_model}.

\noindent From the Friis' equation, the squared magnitudes  of the dependent current sources are given by:
\begin{subequations}
    \begin{align}
    |I_{s_1}(f)|^2 &= 4\,|I_{R_2}(f)|^2\left(\frac{c}{4{\pi}fd}\right)^2G_\text{r}\,G_\text{t}\,\frac{R_2}{R_1}\,\, [\textrm{A}^2], \label{Friss_reaction}\\
    |I_{s_2}(f)|^2 &= 4\,|I_{R_1}(f)|^2\left(\frac{c}{4{\pi}fd}\right)^2G_\text{r}\,G_\text{t}\,\frac{R_1}{R_2}\,\, [\textrm{A}^2].\label{Friis}
    \end{align}
\end{subequations}
From the channel model, the admittance matrix. $\mathbf{Y}_C$, can be calculated using basic circuit analysis, defining $s = j2\pi f$:
\begin{equation}\label{admittance}
\mathbf{Y}_C=\left[\begin{array}{ll}
\frac{(saR_2)^2 + (cR_2)^2 + sacR_2^2 - 4\,\left(\frac{c}{4\pi fd}\right)^2\,sac\,R_2^2\,\sqrt{G_\text{t}\,G_\text{r}}}{((saR_2)^2 + (cR_2)^2 + sacR_2^2)R_1}  & \frac{-2c(sa)^2\,\sqrt{G_\text{t}G_\text{r}}}{4\pi fd ((saR_2)^2 + (cR_2)^2 + sacR_2^2)}\sqrt\frac{R_2^3}{R_1} \\
\frac{-2c\sqrt{G_\textrm{t}G_\textrm{r}}(sa)^2}{4\pi fd ((saR_2)^2 + (cR_2)^2 + sacR_2^2)}\sqrt\frac{R_2^3}{R_1} & \frac{sa(saR_2 + R_2c)}{(saR_2)^2 + cR_2(saR_2 + R_2c)} 
\end{array}\right]\, [\textrm{S}].
\end{equation}
Note here that $\mathbf{Y}_C$ is a symmetric matrix, owing to the reciprocity of antennas. Also, it is common for the signal attenuation between the transmitter and receiver to be extremely large. From the admittance matrix in (\ref{admittance}), it can be verified that, in the far-field region (i.e., sufficiently large distance $d$), $|(\mathbf{Y}_C)_{1,2}| = |(\mathbf{Y}_C)_{2,1}| \ll |(\mathbf{Y}_C)_{1,1}|$. Further, the $(\mathbf{Y}_C)_{1,1}$ entry of the admittance matrix consists of two terms in the numerator with one term being much smaller in magnitude than the other. Consequently, the admittance matrix can be accurately approximated as follows:
\begin{equation}\label{admittance_approx}
\mathbf{Y}_C\approx\left[\begin{array}{ll}
\hspace{2.3cm}\frac{1}{R_1}  & \hspace{1.7cm}0 \\
\frac{-2c\sqrt{G_\textrm{t}G_\textrm{r}}(sa)^2}{4\pi fd ((saR_2)^2 + (cR_2)^2 + sacR_2^2)}\sqrt\frac{R_2^3}{R_1} & \frac{sa(saR_2 + R_2c)}{(saR_2)^2 + cR_2(saR_2 + R_2c)}
\end{array}\right] \, [\textrm{S}] .
\end{equation}
The approximate expression in (\ref{admittance_approx}) is commonly referred to as the unilateral approximation, that is, the electrical properties at the transmit-side antenna ports are independent of what happens at the receiver. This approximation is an almost exact one for far-field wireless communication systems~\cite{ivrlavc2010toward}.

In summary, the model for the antenna channel in (\ref{admittance_approx}) is the circuit model describing the far-field interaction of two line of sight antennas in which the receiving antenna is constrained to have a maximum size of $a$ meters. Such a model is an extension to the circuit theoretic model from \cite{ivrlavc2010toward} that incorporates both the Chu's antenna model and the Bode/Fano matching network.
\section{Computation of the achievable rate}\label{section_4}
In this section, we find the achievable rate in (\ref{capacity_1}) based on the circuit model of SISO communication in Fig.~\ref{fig:two-port_channel} and \ref{fig:channel_model} using the unilateral channel model in (\ref{admittance_approx}). Also observing that the noise current, $I_{\rm{N},1}(f)$, and the Friis' reaction term, $I_{s_1}(f)$, in (\ref{Friss_reaction}) are very small compared to the signal current, they can be ignored as part of the unilateral approximation. The goal is to establish the relationship between the real pass-band voltage waveform, $v(t)$, and output current waveform, $i_{\rm{L}}(t)$, in (\ref{capacity_1}). First, notice that the output current process $i_{\rm{L}}(t)$ can be made Gaussian since the noise sources are Gaussian distributed and all of the circuits are linear. With this, the optimal input voltage process, $v(t)$, must necessarily be a Gaussian random process. We are now ready to state the following result,
\begin{result}
  Let $P_t(f)$, $H(f)$, $\Gamma(f)$ be the continuous transmit power spectral density, the transfer function of the propagation channel, and the network reflection coefficient. If $v(t)$ and $i_L(t)$ are jointly Gaussian random processes, then the mutual information per unit time between the two processes is given by:
  \begin{equation}\label{capacity_russian}
       \int_{0}^ {\infty}\log_2\left(1 + \frac{P_{\rm t}(f)|H(f)|^2(1 - |\Gamma(f)|^2)}{N_0(1 - |\Gamma(f)|^2) + N_{\rm{LNA}}}\right)\mathrm{d}f 
  \end{equation}
  where $N_0 = k_\text{b}\,T$ and $N_{\rm{LNA}} = k_\text{b}\,T\,(N_\text{f} - 1)$.
\end{result}
\begin{proof}
This is a standard result found in most information theory books, e.g. \cite{gallager1968information}, adopted to our particular circuit model of communication. The ratio inside the logarithm corresponds to the signal-to-noise ratio (SNR) between the signal power to the effective noise power as measured at the load $Z_L$.
\end{proof}
To obtain the maximum achievable rate, we need to maximize the mutual information in (\ref{capacity_russian}), over the transmit power spectral density with its constraint in (\ref{transmit_PSD_constraint}), and the reflection coefficient with its two integral constraints in (\ref{constraint_1}) and (\ref{constraint_2}). Notice that since the mutual information is monotonically increasing in the transmit power, then its maximum is achieved by taking $P_{\rm t}(f) = E_{\text{max}}$. If $P_t(f)$ is limited to some bandwidth $\textrm{BW}$, the optimum power spectrum is a constant $E_{\textrm{max}}$ over the bandwidth $\textrm{BW}$ of interest and zero elsewhere. We will use $P^*_t(f)$ for the optimum transmit power spectrum.

The rest of the optimization problem can be formulated by making the substitution, $\mathcal{T}(f) = |T(f)|^2 = 1 - |\Gamma(f)|^2$, thereby leading to:
\begin{eqnarray}\label{capacity}
C_{[b/s]} ~=~ \max_{\mathcal{T}(f), \,\gamma}\, \int_{0}^ {\infty}\log_2\left(1 + \frac{P^*_t(f)\,|H(f)|^2\,\mathcal{T}(f)}{N_0\,\mathcal{T}(f) + N_{\rm{LNA}}}\right)\textrm{d}f 
\end{eqnarray}

\begin{eqnarray}\label{signum_function}
\textrm{subject to} \begin{cases}
~ \int_{0}^{\infty} f^{-2} \ln \left(\frac{1}{1 - \mathcal{T}(f)}\right)\, \mathrm{d}f=2\pi^2\left(\frac{2 a}{c}-2 \gamma^{-1}\right) \triangleq K_1\\
~ \int_{0}^{\infty} f^{-4} \ln \left(\frac{1}{1-\mathcal{T}(f)}\right)\, \mathrm{d}f=8\pi^4\left(\frac{4 a^{3}}{3 c^{3}}+\frac{2}{3} \gamma^{-3}\right) \triangleq K_2\\
~  0 \leq\mathcal{T}(f) \leq 1, ~\forall f.
\end{cases}
\end{eqnarray}
The Lagrangian associated with (\ref{capacity}) and (\ref{signum_function}) is given by
\begin{equation}\label{eq:augmented-lagragian}
    \begin{aligned}[b]
        J(\mathcal{T}(f)) &= \int_{0}^ {\infty}\log_2\left(1 + \frac{P^*_t(f)\,|H(f)|^2\,\mathcal{T}(f)}{N_0\,\mathcal{T}(f) + N_{LNA}}\right)\textrm{d}f \\
        &\hspace{1cm}+ \mu_1\left(\int_{0}^{\infty} f^{-2} \ln\bigg( \frac{1}{1 - \mathcal{T}(f)}\bigg) \mathrm{d}f - K_1\right) \\
        &\hspace{2cm}+ \mu_2\left(\int_{0}^{\infty} f^{-4} \ln\bigg( \frac{1}{1-\mathcal{T}(f)}\bigg) \mathrm{d}f - K_2\right) \\
        &\hspace{3cm}+ \mu_3\,\mathcal{T}(f) + \mu_4\,(\mathcal{T}(f)-1).
    \end{aligned}
\end{equation}
From the gradient condition of variational calculus and complimentary slackness, it follows that:
\begin{equation}\label{variational_calculus}
    \frac{\textrm{d}}{\textrm{d}\varepsilon}\big[J(\mathcal{T}^{\tiny{\starletfill}}(f) + \varepsilon\,\eta(f))\big]\Bigg|_{\varepsilon = 0} =~ 0, 
\end{equation}
for all functions $\eta(f)$ (see Appendix \ref{appendix:calculus-of-variation} for more details). By taking the derivative in (\ref{variational_calculus}) and solving for $\mathcal{T}^{\tiny{\starletfill}}(f)$, the optimal reflection coefficient is given by the solution of the quadratic equation. By showing that $\mathcal{T}^{\tiny{\starletfill}}(f) \leq 1$ (see the proof in Appendix \ref{appendix3:derivation-optimal-transmission-coeff}), we obtain 
\begin{equation}\label{optimal_transmission}
    \mathcal{T}^{\tiny{\starletfill}}(f) ~=~ \max \Bigg(0,\frac{-C_2(f) - \sqrt{C_2^2(f) - 4\,C_1(f)\,C_3(f)}}{2\,C_1(f)}\Bigg),
\end{equation}
where 
\begin{subequations}\label{quadratic solution}
    \begin{align}
    &C_1(f) = (N_0 + P^*_t(f)|H(f)|^2)\,N_0\,(\mu_1\,f^{-2} + \mu_2\,f^{-4}), \\
    &C_2(f) = (2\,N_0\,N_{\textrm{LNA}} + N_{\rm LNA}\,P^*_t(f)|H(f)|^2)\,(\mu_1\,f^{-2} + \mu_2\,f^{-4}) - P^*_t(f)|H(f)|^2\,N_{\textrm{LNA}}, \\
    &C_3(f) = P^*_t(f)|H(f)|^2\,N_{\textrm{LNA}} + N_{\textrm{LNA}}^2\,(\mu_1\,f^{-2} + \mu_2\,f^{-4}).
    \end{align}
\end{subequations}
Maximizing the Lagrangian w.r.t. $\gamma$ (see (\ref{appendix:optimal-gamma}) in Appendix \ref{appendix3:derivation-optimal-transmission-coeff}), the optimum is given by:
\begin{equation}
    \gamma = 2\pi\sqrt{\frac{\mu_2}{\mu_1}}.
\end{equation}
Solving the first constraint in (\ref{signum_function}) for $a$, we express the antenna size as function of the Lagrange multipliers $\mu_1$ and $\mu_2$, as follows:
\begin{equation}\label{antenna_size_eqn}
    a(\mu_1,\mu_2) ~=~ \frac{c}{4\pi^2} \int_{0}^{\infty} f^{-2} \ln \Bigg(\frac{1}{1 - \mathcal{T}^{\tiny{\starletfill}}(f)}\Bigg) \,\mathrm{d}f + \frac{c}{2\pi}\sqrt{\frac{\mu_1}{\mu_2}}~[\textrm{m}].
\end{equation}
From the second constraint, we can obtain an implicit relation between the Lagrange multipliers $\mu_1$ and $\mu_2$:
\begin{equation}\label{constraint_eqn}
    \frac{1}{8\pi^4} \int_{0}^{\infty} f^{-4} \ln\Bigg(\frac{1}{1-\mathcal{T}^{\tiny{\starletfill}}(f)}\Bigg) \mathrm{d}f=\frac{4 a\,(\mu_1,\mu_2)^{3}}{3 c^{3}}+ \frac{1}{12\pi^3}\left(\frac{\mu_1}{\mu_2}\right)^{3/2}.
\end{equation}
Besides, from (\ref{constraint_eqn}), we can numerically solve for $\mu_2$ for fixed $\mu_1$ which means we need only to specify $\mu_1$ and obtain the antenna size from (\ref{antenna_size_eqn}). This is in fact equivalent to fixing the size as well as $\gamma$ and obtaining the corresponding Lagrange multipliers $\mu_1$ and $\mu_2$.

We now summarize in Result~\ref{result2:summary}, the main optimization steps obtained in this section.
\begin{result}\label{result2:summary}
For a given size of the receiver antenna structure, $a$, the matched electrical Chu's antenna circuit has the following
    \begin{enumerate}
       \item An optimal reflection coefficient [cf. (\ref{optimal_transmission})]
            \begin{equation*}
                \mathcal{T}^{\tiny{\starletfill}}(f) ~=~ \max \Bigg(0,\frac{-C_2(f) - \sqrt{C_2^2(f) - 4\,C_1(f)\,C_3(f)}}{2C_1}\Bigg),
            \end{equation*}
            where $C_1(f)$, $C_2(f)$ and $C_3(f)$ are defined in (\ref{quadratic solution}).
    \item An achievable rate equal to [cf. (\ref{capacity})]
      \begin{eqnarray}\label{eqn:mutual_info}
            C_{[b/s]} = \int_{0}^ {\infty}\log_2\left(1 + \frac{P^*_t(f)\,|H(f)|^2\,\mathcal{T}^{\tiny{\starletfill}}(f)}{N_0\,\mathcal{T}^{\tiny{\starletfill}}(f) + N_{\rm{LNA}}}\right)\,\textrm{d}f .
        \end{eqnarray}
    \item Two constraints (\ref{antenna_size_eqn}) and (\ref{constraint_eqn}) involving the antenna size, $a(\mu_1,\mu_2)$, as a function of the two Lagrange multipliers $\mu_1$ and $\mu_2$:
            \begin{equation*}
            \begin{aligned}
                &a(\mu_1,\mu_2) = \frac{c}{4\pi^2} \int_{0}^{\infty} f^{-2} \ln \Bigg(\frac{1}{1 - \mathcal{T}^{\tiny{\starletfill}}(f)}\Bigg) \mathrm{d}f + \frac{c}{2\pi}\sqrt{\frac{\mu_1}{\mu_2}}~[\textrm{m}],\\
        &\frac{1}{8\pi^4} \int_{0}^{\infty} f^{-4} \ln\Bigg( \frac{1}{1-\mathcal{T}^{\tiny{\starletfill}}(f)}\Bigg) \mathrm{d}f=\frac{4 a(\mu_1,\mu_2)^{3}}{3 c^{3}}+ \frac{1}{12\pi^3}\left(\frac{\mu_1}{\mu_2}\right)^{3/2}.
            \end{aligned}
            \end{equation*}
    \end{enumerate}
\end{result}
This result gives the maximum mutual information of a physically realizable antenna of some fixed size $a$. In fact, any other antenna structure of the same size would be able to achieve mutual information that is strictly smaller than what is given in (\ref{eqn:mutual_info}). It is important to briefly examine the synthesis of the optimal matching network obtained via variational optimization. From the shape of the reflection coefficient, which is not constant over the pass-band of interest, the well-known synthesis techniques based on Butterworth, Chebyshev, or elliptic filters could not be utilized. In general, the best approximation to the obtained optimal matching network by means of a reactive ladder structure of a given order, $N$, is an important problem for future investigation. 
 In the next section, we consider a different setting where the contribution of the interference term to the received power is taken into account.

\section{Analysis of interference}\label{sec:interference}

In this section, we extend our model to the homogeneous interference scenario. From the perspective of the equivalent circuit model of communication in Fig.~\ref{fig:channel_model}, the extension is made to incorporate an additional current source to model the effect of interference. Such an approach is probably the simplest physically consistent modeling of interference. To produce an analytically tractable model, we leverage tools from stochastic geometry to derive the first two moments of the interference power density. This approach does not require complex computations to estimate the distributions' parameters \cite{heath2013modeling, akoum2013interference}.

We consider a downlink cellular system where the source of interference is a single type of interferer, namely a macrocell of radius $R_0$. In this model, the interferer's locations $x_i \in \mathbb{R}^2$ and small scale powers $p_i \in \mathbb{R}^+$ follow a 2-dimensional Marked Point Process $\Phi = \{x_{i}, p_i\}$. The marks, $p_i \in \mathbb{R}^+$, correspond to the squared magnitude of the normalized  small-scale channel parameter and follow a unit-mean distribution $\mathbb{P}[\,p_i \leq s\,]=G(s)$. Additionally, we consider a Poisson Point Process (PPP) with density $\rho$ such that the marks are exponentially distributed according to the small scale power distribution of the Rayleigh faded channel, i.e. $G(s)=1 - e^{-s}$. Finally, we also consider an ommi-directional path-loss (OPL) function $l(r) = (\frac{r}{\lambda})^{\alpha}$ where $\alpha$ is called the path-loss exponent \cite{Baccelli} and guarantees the finiteness of the total interference power, i.e., $\alpha > 2$.
\subsection{Gamma $2^{nd}$ Order Moment Matching}\label{subsec:gamma-matching}
Under the model assumptions stated above, the mean and variance of the total received interference power $I$ with Rayleigh fading interference channels are well-known to have the following expressions [cf. \cite{Baccelli}].
\begin{subequations}
    \begin{align}
    \mathbb{E}[I]& = \frac{2 \pi \,\rho}{\alpha-2} \,P_{\rm t}\,\lambda^{\alpha}\,R_0^{2-\alpha}~[\textrm{W}],\label{eq:mean-total-received-power}\\
    \sigma_{I}^2 
\triangleq \textrm{Var}[I] &= 2\,P_{\rm t}^{2}\,\frac{\pi \,\rho}{\alpha-1} \,\lambda^{2\alpha}\,R_0^{2(1-\alpha)}\,[\textrm{W}^2],
\label{eq:variance-total-received-power}
    \end{align}
\end{subequations}
Now, given the finite first $\mathbb{E}[I]$ and second order $\mathbb{E}[I^2]$ moments of the interference power $I$ from (\ref{eq:mean-total-received-power}) and (\ref{eq:variance-total-received-power}), the characterization of the distribution of the interference power $I$, $p_{\textrm{I}}(I)$, can be approximately achieved by using the second-order moment matching with the Gamma distribution \cite{heath2011multiuser}.
The parameters $k$ and $\theta$ of the matched Gamma distribution $q_{I}(I) = \Gamma(I;k,\theta)$ of the interference power $I$ are explicitly given by:\\
\begin{subequations}
  \label{eq:k-theta}
  \begin{tabularx}{\textwidth}{Xp{0cm}X}
  \begin{equation}
     k = \frac{\mathbb{E}[I]^2}{\sigma_{I}^2} = 2\pi\,\rho\,R_0^2\,\frac{(\alpha-1)}{(\alpha-2)^2},
  \end{equation}
  & &
  \begin{equation}
   \theta = \frac{\sigma_{I}^2}{\mathbb{E}[I]} = \frac{(\alpha-2)}{(\alpha-1)}\,P_{\rm t}\,\Big(\frac{\lambda}{R_0}\Big)^{\alpha}.
  \end{equation}
  \end{tabularx}
\end{subequations}
This approximation not only yields a tractable interference distribution but also avoids the need for the Laplace characterizations of (\ref{eq:mean-total-received-power}) and (\ref{eq:variance-total-received-power}).

From a circuit perspective, the fact that the interference field of a set of transmitters can be interpreted as a noise field allows us to extend the interference-free model of Fig. \ref{fig:channel_model} to handle the interference case. For this reason, we treat the interference term, similarly to the noise term, as an additional independent current source $I_{\textrm{inter}}(f)$ in parallel to the noise current source $I_{s_2}(f)$.
\subsection{Computation of achievable rate}
When the interference $I$ is taken into account, the mutual information (\ref{capacity}) with an optimal matching network needs to be slightly modified by augmenting the environmental noise power $N_0$ with interference power $I$ (i.e $N_0 \Rightarrow N_0 + I$ in (\ref{quadratic solution})). In presence of interference, the optimal matching network depends on the realization of the random interference power $I$. 
By averaging the mutual information w.r.t. $p_{\textrm{I}}(I)$, we find that although the average mutual information does not admit a closed-form expression in the general case, it can be found easily numerically. It is worth noting, however, that it admits an approximate closed-form expression only when the transmission coefficient $\mathcal{T}(f)$ is not a function of interference (i.e., a fixed antenna with a matching network structure that cannot be adapted to interference). Notice that, in presence of interference, an optimal Chu's antenna must necessarily be reconfigurable. In other words, the receive antenna should have the capability to adjust its internal matching network circuitry such that the transmission coefficient is maximized for every interference realization value. In the sequel, we distinguish two cases.
\subsubsection{Mutual information with fixed antenna}
By averaging over the matched interference density $q(I)$, we rewrite the mutual information (\ref{capacity}), in this case, as follows
\begin{equation}
\begin{aligned}[b]\label{eq:capacity-no-matching}
    C_{[b/s]} &= \mathbb{E}_{q(I)}\Bigg[\int_{0}^ {\infty}\log_2\left(1 + \frac{P^*_t(f)\,|H(f)|^2\,\mathcal{T}(f)}{(N_0 + I)\,\mathcal{T}(f) + N_{\rm{LNA}}}\right)\textrm{d}f\Bigg]\\
    &= \int_{0}^ {\infty}\mathbb{E}_{q(I)}\Bigg[\log_2\left(1 + \frac{P^*_t(f)\,|H(f)|^2\,\mathcal{T}(f)}{(N_0 + I)\,\mathcal{T}(f)+ N_{\rm{LNA}}}\right)\Bigg]\, \textrm{d}f.
    \end{aligned}
\end{equation}
where the transmission coefficient, $\mathcal{T}(f)$, does not depend on the interference power $I$. In the special case where the impedance matching is not taken into account, one can set $\mathcal{T}(f)$ to $1-|\widetilde{\Gamma}(f)|^2$ with $\widetilde{\Gamma}(f)$ being the reflection coefficient without matching (\ref{eq:no-matching-reflection-coeff}) established in Appendix \ref{Fano_appendix}.\\
To derive an approximate closed form to the integrand of (\ref{eq:capacity-no-matching}), we start from the fact that
\begin{equation}\label{eq:sum-gamma-variables}\small
    \begin{aligned}[b]
        \mathbb{E}_{q(I)}\Bigg[\log_2\left(1 + \frac{P^*_t(f)\,|H(f)|^2\,\mathcal{T}(f)}{(N_0 + I)\,\mathcal{T}(f) + N_{\rm{LNA}}}\right)\Bigg]&=\mathbb{E}_{q(I)}\Bigg[\log_2\Big( \big(I+N_0 +P^*_t(f)\,|H(f)|^2\big) \,\mathcal{T}(f)+ N_{\rm{LNA}}\Big)\Bigg]\\
        &\quad~~-\mathbb{E}_{q(I)}\Bigg[\log_2\Big( (I+N_0) \,\mathcal{T}(f)+ N_{\rm{LNA}}\Big)\Bigg].
    \end{aligned}
\end{equation}
By applying the second order Taylor expansion of $f: x \rightarrow \mathbb{E}[\log_2(1+x)]$ around $\mathbb{E}[x]$, i.e.,
\begin{equation*}\label{eq:approx-E[log(1+x)]}
    \mathbb{E}[\log_2 (1+x)] = \log_2 (1+\mathbb{E}[x])-\frac{\operatorname{Var}[x]}{2\,(1+\mathbb{E}[x])^{2}} + o(\operatorname{Var}[x]),
\end{equation*}
to the two terms of the right-hand side of (\ref{eq:sum-gamma-variables}) separately, we get the following result:

\begin{result}\label{result:summary}In the homogeneous interference scenario, the approximate closed-form expression of the average mutual information when the transmission coefficient $\mathcal{T}(f)$ is independent of interference, is given by:
\begin{equation}\label{eq:closed-form-approx-expectation-log}
  \begin{aligned}[b]
        C_{[b/s]} &= \mathbb{E}_{q(I)}\Bigg[\log_2\left(1 + \frac{P^*_t(f)\,|H(f)|^2\,\mathcal{T}(f)}{(N_0 + I)\,\mathcal{T}(f) + N_{\rm{LNA}}}\right)\Bigg] \\
        &= \log_2\Big( \big((\mathbb{E}[I] + N_0 + P^*_t(f)\,|H(f)|^2) \,\mathcal{T}(f)+ N_{\rm{LNA}}\big)/\big((\mathbb{E}[I] + N_0) \,\mathcal{T}(f) + N_{\rm{LNA}}\big)\Big)\\
        &\hspace{1cm}- \frac{\mathcal{T}(f)^2\,\sigma_{I}^2}{2}\Big(\big((\mathbb{E}[I] + N_0+P^*_t(f)\,|H(f)|^2) \,\mathcal{T}(f)+ N_{\rm{LNA}}\big)^{-2}\\ &\hspace{2.5cm}-\big((\mathbb{E}[I] + N_0) \,\mathcal{T}(f)+ N_{\rm{LNA}} \big)^{-2}\Big) +o(\sigma_I^2),
    \end{aligned}
\end{equation}
where $\mathbb{E}[I]$ and $\sigma_{I}^2$ are the mean and variance of $q(I)$ obtained from (\ref{eq:mean-total-received-power}) and (\ref{eq:variance-total-received-power}), respectively. The accuracy of such approximation is high when the last correction term in (\ref{eq:closed-form-approx-expectation-log}) is much smaller than 1, which corresponds to the condition $\sigma_I^2 \ll \mathbb{E}[I]^2$. The case where the pathloss exponent $\alpha$ is close to 2 would satisfy this requirement.
\end{result}
\subsubsection{Mutual information with adaptive antenna and an optimal matching network}
The mutual information (\ref{capacity}) is averaged over its matched interference density $q(I)$, i.e.,
\begin{equation}\label{eq:capacity-with-interference}
    C_{[b/s]} = \mathbb{E}_{q(I)}\Bigg[\int_{0}^ {\infty}\log_2\left(1 + \frac{P^*_t(f)\,|H(f)|^2\,\mathcal{T}^{\tiny{\starletfill}}(f)}{(N_0 + I)\,\mathcal{T}^{\tiny{\starletfill}}(f) + N_{\rm{LNA}}}\right)\textrm{d}f\Bigg].
\end{equation}
As was mentioned earlier, the mutual information in  (\ref{eq:capacity-with-interference}) cannot be found in closed form. 
In this case, we only compute it numerically (\ref{eq:capacity-with-interference}) in Section \ref{section_6}.

Now that we derived the expression of the antenna mutual information under the size constraint with and without considering the interference, we are ready to compare it to the standard size-unconstrained Shannon limit and examine the effect of the optimal impedance matching.
\section{Numerical Results and Discussion}\label{section_6}
The findings in this section are based on the numerical evaluation of the following expressions
\begin{enumerate}
    \item \textit{Interference-free scenario}: the achievable rate $C_a$ (\ref{capacity}) as well as antenna size (\ref{antenna_size_eqn}) and physical realizability constraint in (\ref{constraint_eqn}). The transmission coefficients are given in (\ref{optimal_transmission}) and (\ref{transmission_nomatch}) with optimal matching and no matching, respectively.
    \item \textit{Interference scenario}: Under the matched density $q(I)$ of the interference power, we consider the two separate cases:
    \begin{itemize}
        \item no matching setting: the approximation (\ref{eq:closed-form-approx-expectation-log}) of the achievable rate $C^{\textrm{no MN}}_{\textrm{inter}}$ (\ref{eq:capacity-no-matching}),
        \item impedance matching setting: the achievable rate  $C^{\textrm{MN}}_{\textrm{inter}}$ (\ref{eq:capacity-with-interference}).
    \end{itemize}
\end{enumerate}
\subsection{Interference-free scenario}
\subsubsection{Simulation results of the SNR}
\begin{figure}
\begin{subfigure}{.5\textwidth}
  \centering
  \includegraphics[width=.75\linewidth]{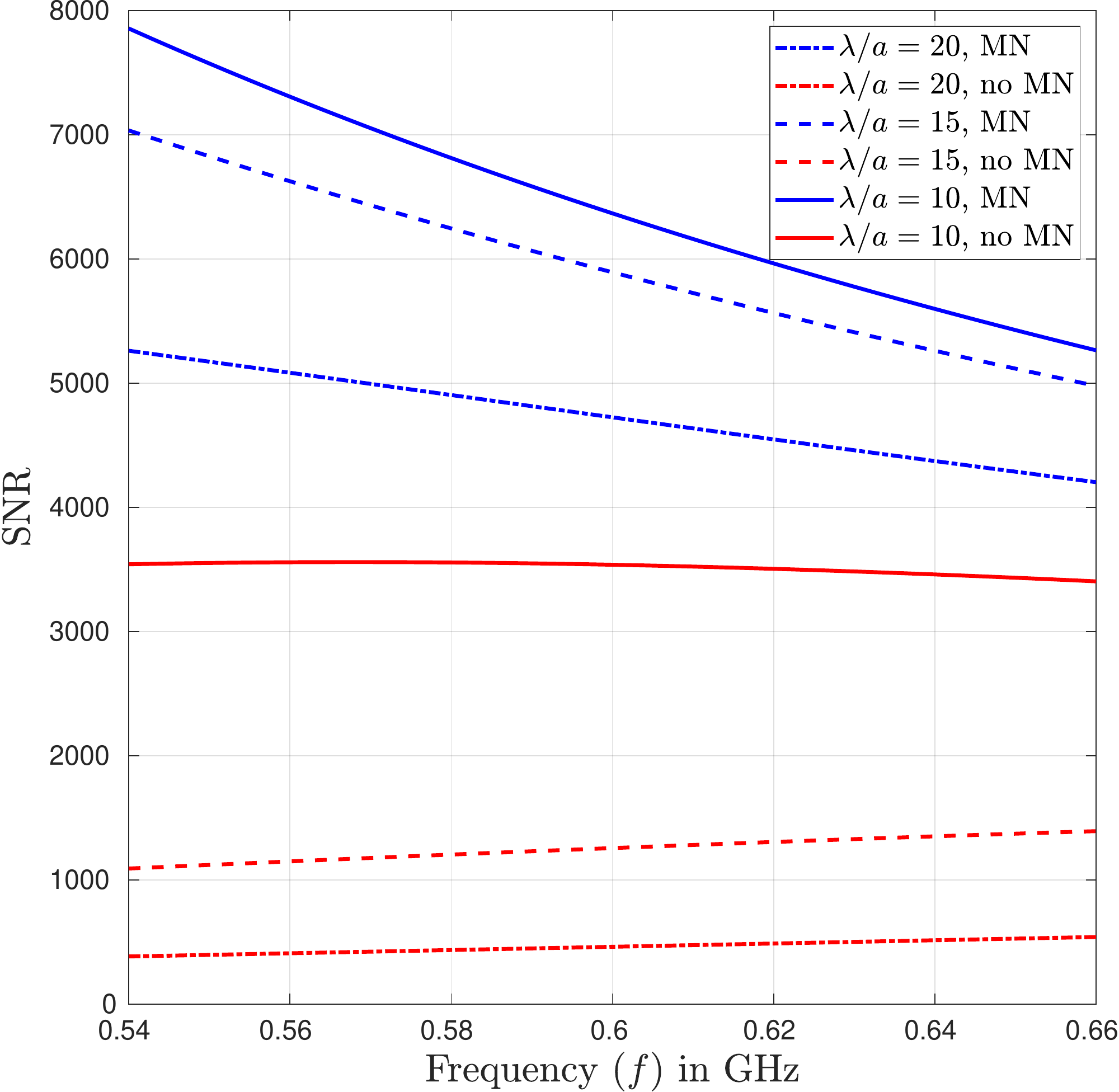}
  \caption{\textrm{SNR} as a function of frequency with $f_c = 600\,[\textrm{MHz}]$, \\\hspace{2cm}$BW = 0.2\,f_c$ and $P = 4\,[W]$.}
  \label{fig:low_frequency_SNR}
\end{subfigure}%
\begin{subfigure}{.5\textwidth}
  \centering
  \includegraphics[width=.75\linewidth]{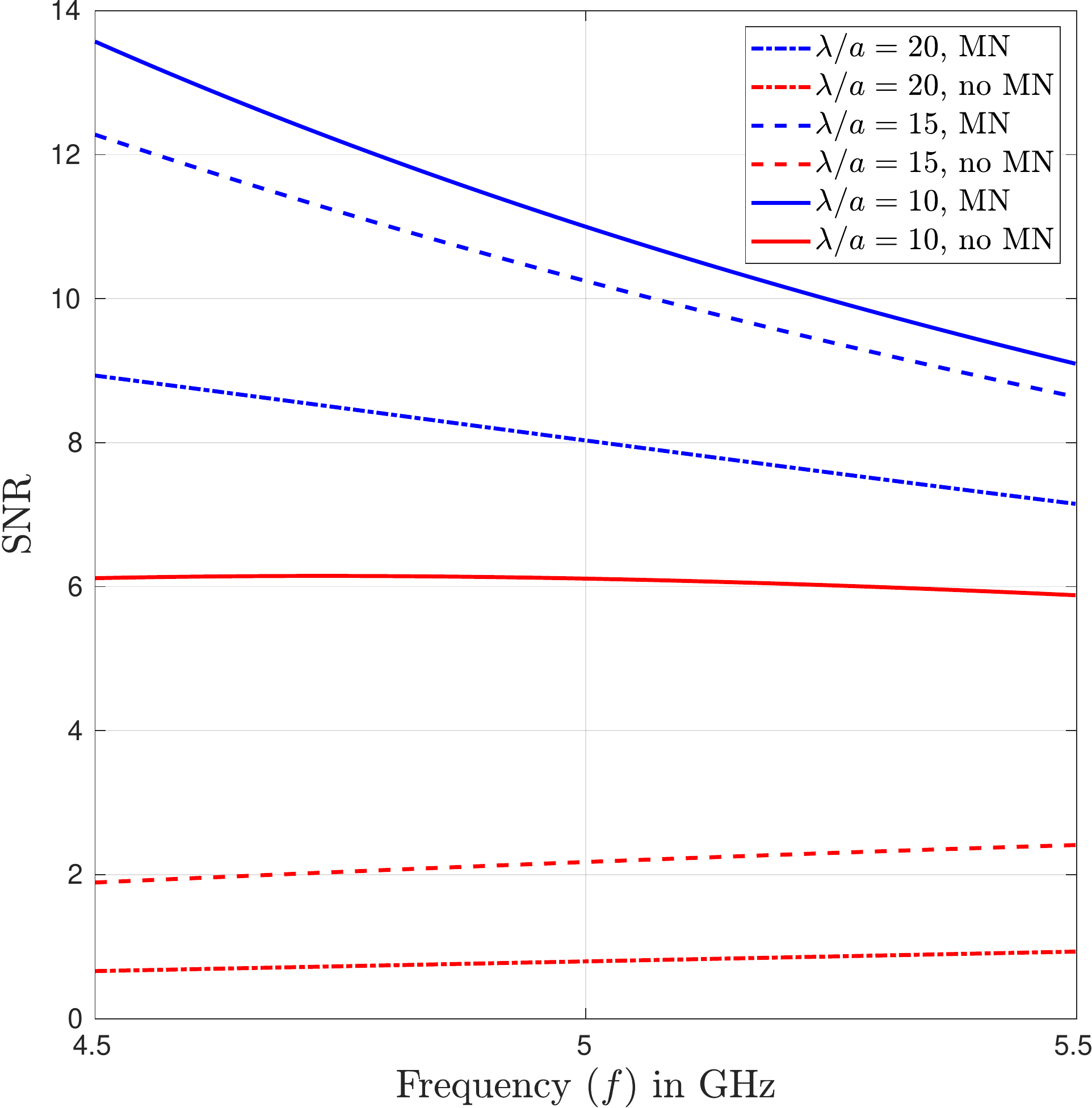}
  \caption{\textrm{SNR} as a function of frequency with $f_c = 5\,[\textrm{GHz}]$, \\\hspace{2cm}$BW = 0.2\,f_c$ and $P = 4\,[W]$.}
  \label{fig:mid_frequency_SNR}
\end{subfigure}
\begin{subfigure}{.5\textwidth}
  \centering
  \includegraphics[width=.75\linewidth]{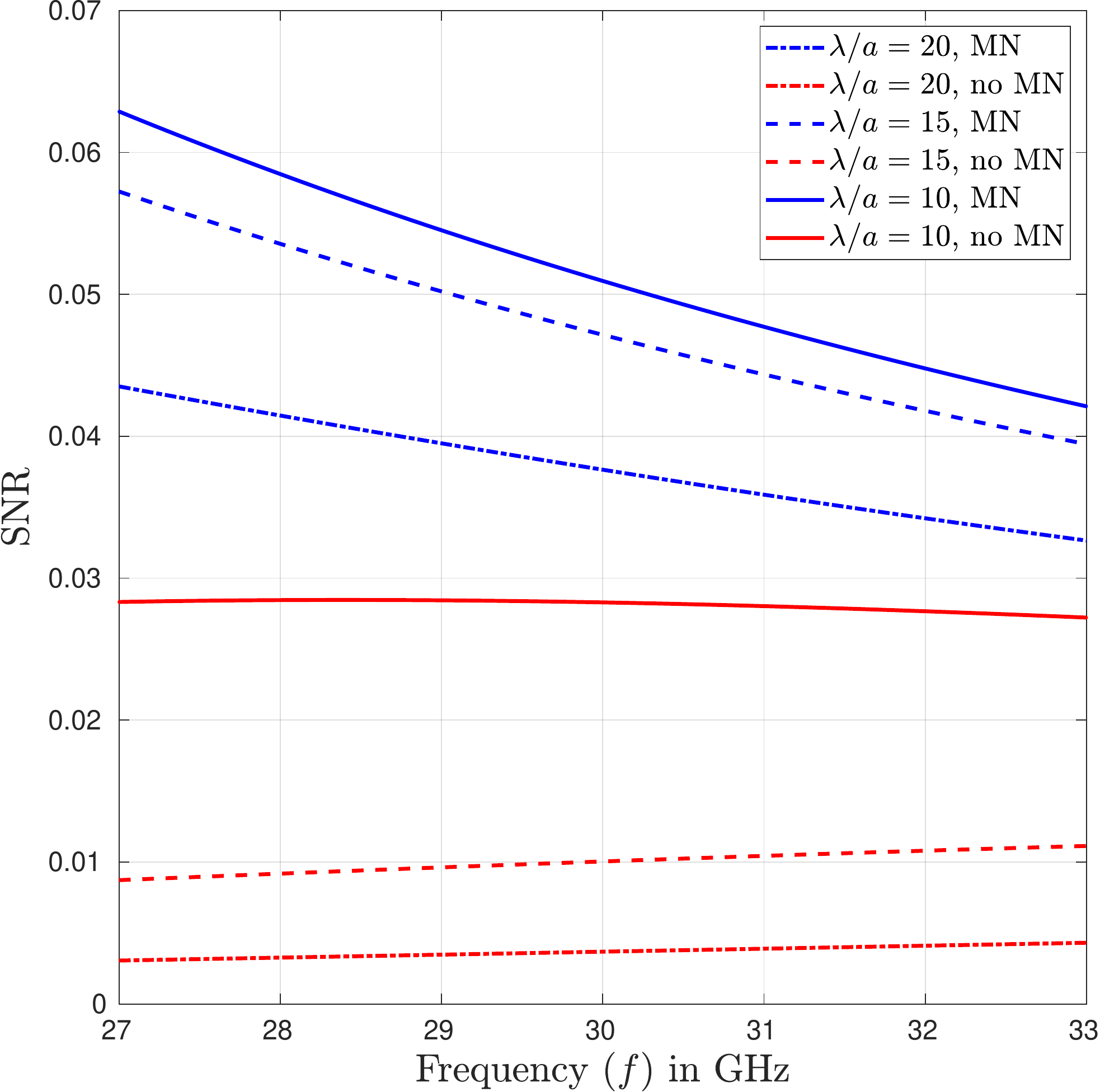}
  \caption{\textrm{SNR} as a function of frequency with $f_c = 30\,[\textrm{GHz}]$, \\\hspace{2cm}$BW = 0.2\,f_c$ and $P = 4\,[W]$.}
  \label{fig:high_frequency_SNR}
\end{subfigure}
\begin{subfigure}{.5\textwidth}
  \centering
  \includegraphics[width=.75\linewidth]{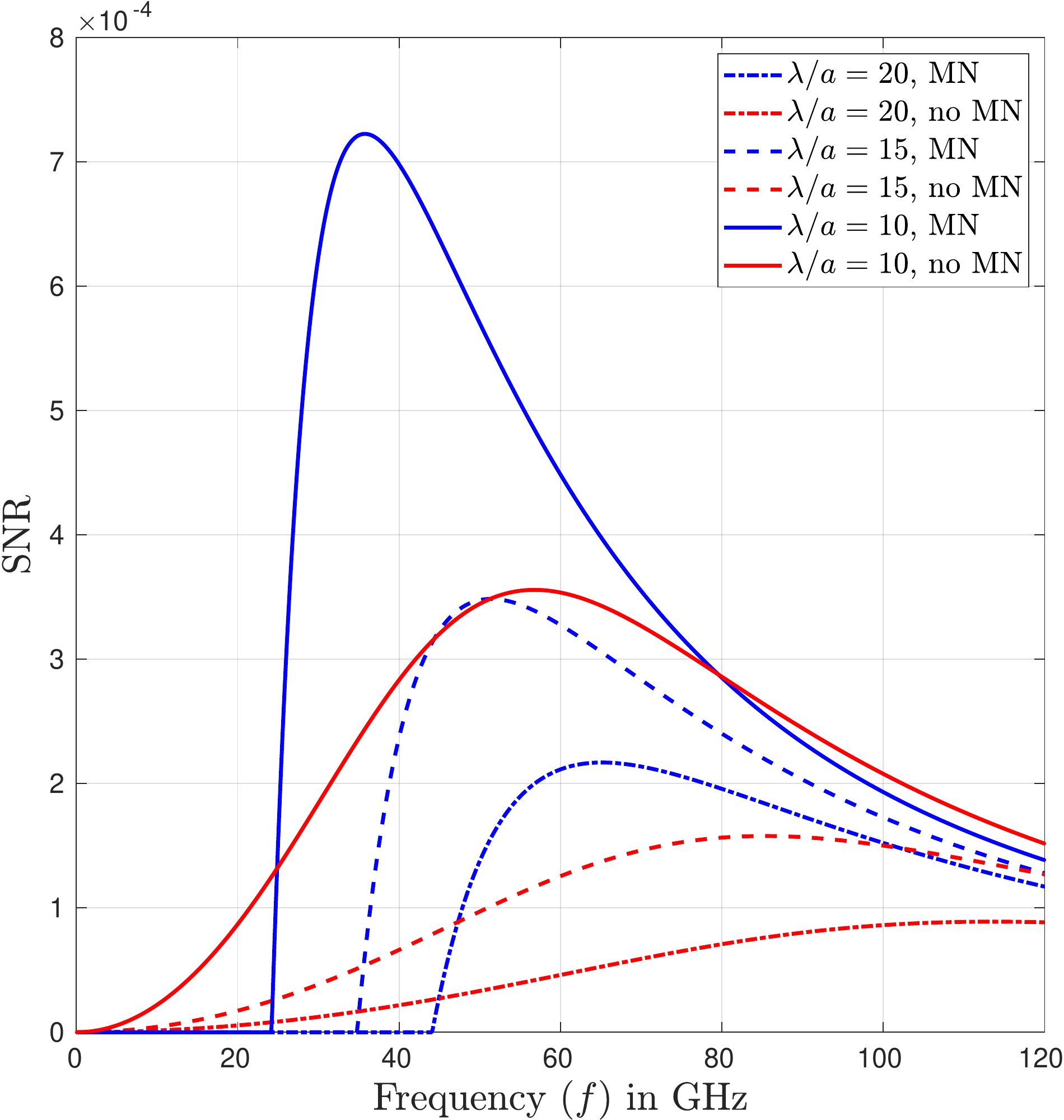}
  \caption{\textrm{SNR} as a function of frequency with $f_c = 60\,[\textrm{GHz}]$,\\\hspace{2cm}$BW = 120\,[\textrm{GHz}]$ and $P = 4\,[W]$.}
  \label{fig:ultrawideband_SNR}
\end{subfigure}
\caption{Plots of $\textrm{SNR}$ as a function of frequency for three different antenna sizes and four different regimes of operation depending on the carrier }
\label{fig:SNR_4_regimes}
\end{figure}

We first compare in Fig.~\ref{fig:SNR_4_regimes} the value of the SNR as a function of the frequency:
\begin{equation}
    \textrm{SNR}(f) ~=~ \frac{P^*_t(f)
    \,|H(f)|^2\,\mathcal{T}(f)}{N_0\,\mathcal{T}(f) + N_{\textrm{LNA}}},
\end{equation}
for three different wavelength to antenna size ratios $\lambda/a \in \{20,15,10\}$ with both the optimal matching network and no matching network by fixing the transmit power to $P=4~[\textrm{W}]$. The distance between the transmitter and receiver is set to $d=1000\,[\textrm{m}]$ and the bandwidth to $\textrm{BW} = 0.2f_c$. Referring back to (\ref{capacity_russian}), we set the noise factor to $N_f = 2$ (or equivalently $3$ dB). The noise temperature $T = 300\,[\textrm{K}]$  such that $N_0 = N_{\rm{LNA}} = k_\text{b}\,T = 4.14\times10^{-21}~[\textrm{J}]$  and $G_t = G_r = 3/2$ (TM$_1$ mode). In Fig.~\ref{fig:SNR_4_regimes}, we explore $4$ different regimes of operation: the low-frequency regime with $f_c = 600\,[\textrm{MHz}]$ in Fig. \ref{fig:low_frequency_SNR}, the medium-frequency regime with $f_c = 5\,[\textrm{GHz}]$ in Fig. \ref{fig:mid_frequency_SNR}, the high-frequency regime with $f_c = 30\,[\textrm{GHz}]$ in Fig. \ref{fig:high_frequency_SNR}, and the ultrawideband-frequency regime with $f_c = 60\,[\textrm{GHz}]$ and $\textrm{BW}=120\,[\textrm{GHz}]$ in Fig. \ref{fig:ultrawideband_SNR}. Unlike Figs. \ref{fig:low_frequency_SNR}, \ref{fig:mid_frequency_SNR} and \ref{fig:high_frequency_SNR} where the bandwidth is equal to $0.2\,f_c$, we increased in Fig. \ref{fig:ultrawideband_SNR} the bandwidth from $0.2\,f_c$ to $2\,f_c$ to show the whole SNR profile across a wider bandwidth and notice the cut of lower frequencies corresponding to a higher value of the reflection coefficient. We observe that the optimal matching network can significantly improve the $\textrm{SNR}$ over all considered frequency regimes as well as over very large bandwidths. The improvement is most evident for the small antenna size, $\lambda/a = 20$, where it is seen that incorporating the MN is more advantageous than doubling the antenna size. We also note that the antenna size $a$ varies significantly from as large as $a = 5\,[\textrm{cm}]$ in the low-frequency regime to as small as $a = 0.25\,[\textrm{mm}]$ in the ultrawideband-frequency regime, thereby confirming that the absolute antenna size $a$ is meaningful w.r.t. the carrier wavelength only.
With no matching and for all considered frequencies, a huge increase in the $\textrm{SNR}$ is observed per two-fold increase in antenna size in the compact antenna regime (SNR improvement for $a \ll \lambda$  can be up to sixteen-fold by doubling the antenna size since transmission coefficient $|\widetilde{T}(f)|\propto a^4/\lambda^4$ in this regime according to (\ref{transmission_nomatch})). Little attention to this bottleneck in the design of communication systems has been given by the communication community, in contrast to the greater importance in the research direction of the antenna design community \cite{hansen2011small}. This is why the circuit/information-theoretic modeling, design, and optimization of communication systems is essential to make antenna theory/design concepts accessible to communication engineers \cite{ivrlavc2010toward}.

\subsubsection{Simulation results of the fraction of the achievable Shannon capacity}
\begin{figure}[h!]
    \centering
    \includegraphics[scale=0.5]{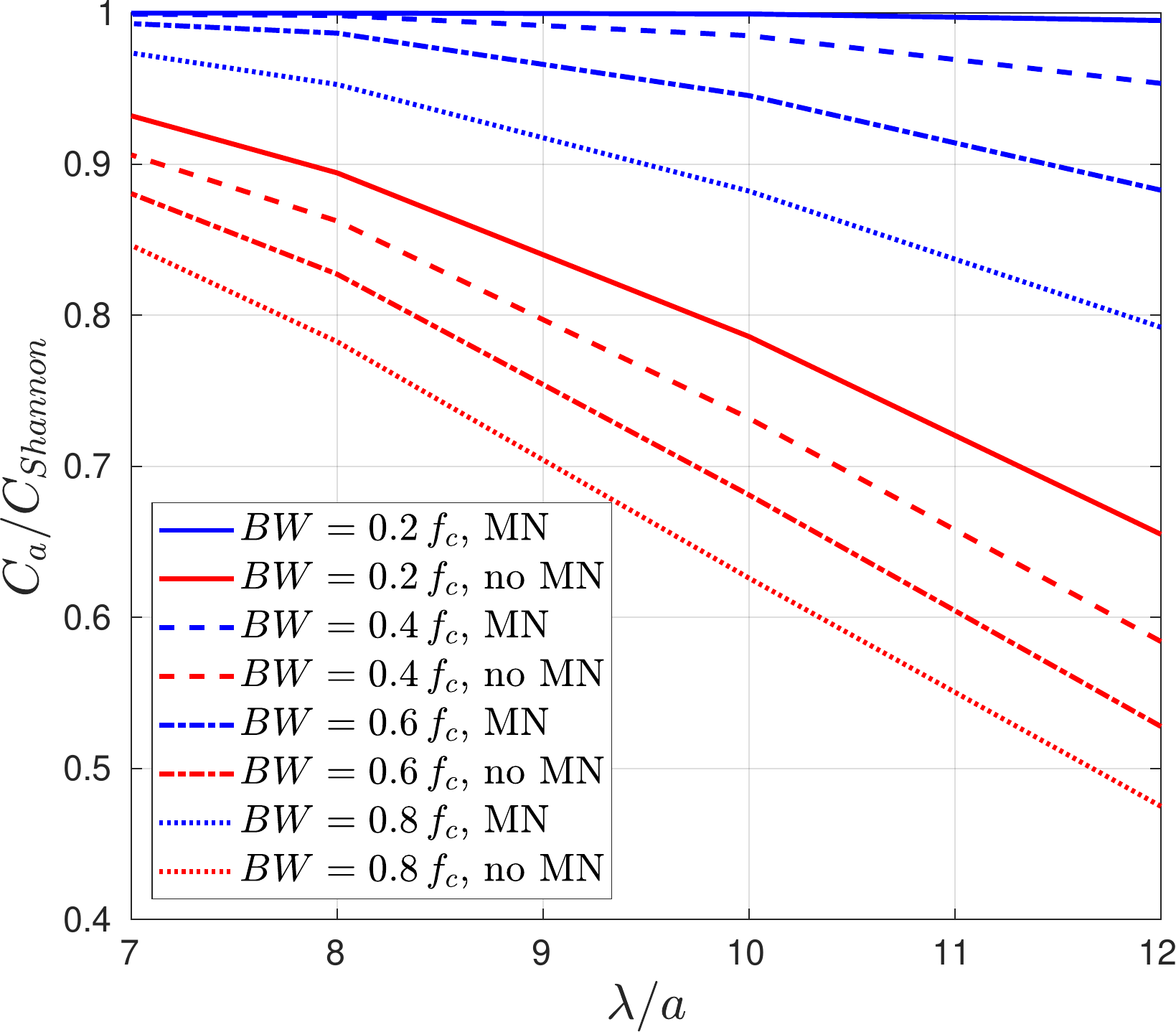}
    \caption{Plots of the fraction of the Shannon capacity that could be achieved with a given size, $a$, as a function of $\lambda/a$ at $f_c = 5\,[\textrm{GHz}]$, $P=4\,[W]$ in four different regimes of operation in terms of bandwidth.}
    \label{fig:antenna_size_capacity}
\end{figure}

\begin{figure}[h!]
\begin{subfigure}{.5\textwidth}
  \centering
  \includegraphics[width=.65\linewidth]{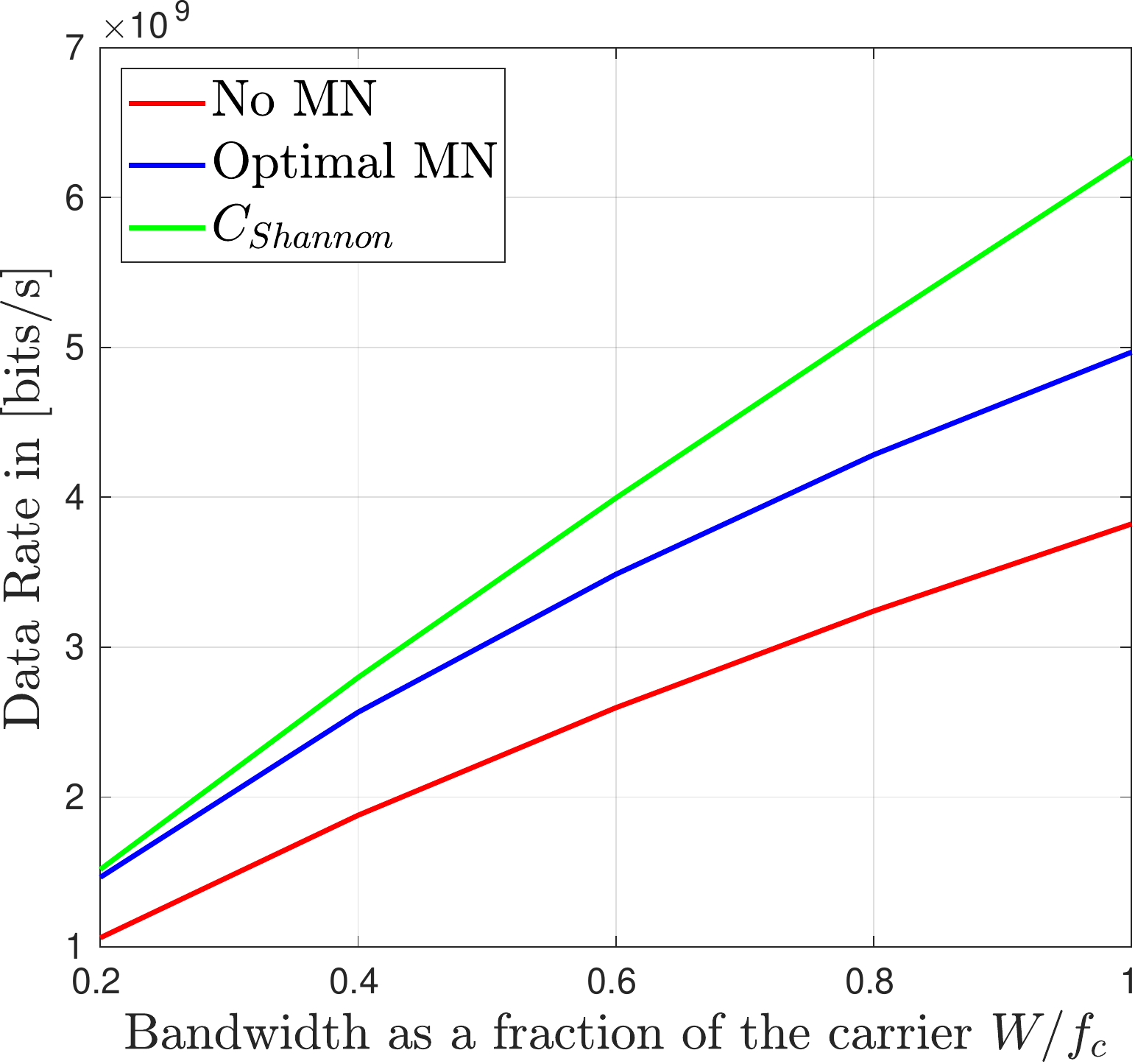}
  \caption{Low-frequency regime: $f_c = 600\,[\textrm{GHz}]$,\,$P=4\,[W]$.}
  \label{fig:Capacity_bandwidth_1}
\end{subfigure}%
\begin{subfigure}{.5\textwidth}
  \centering
  \includegraphics[width=.65\linewidth]{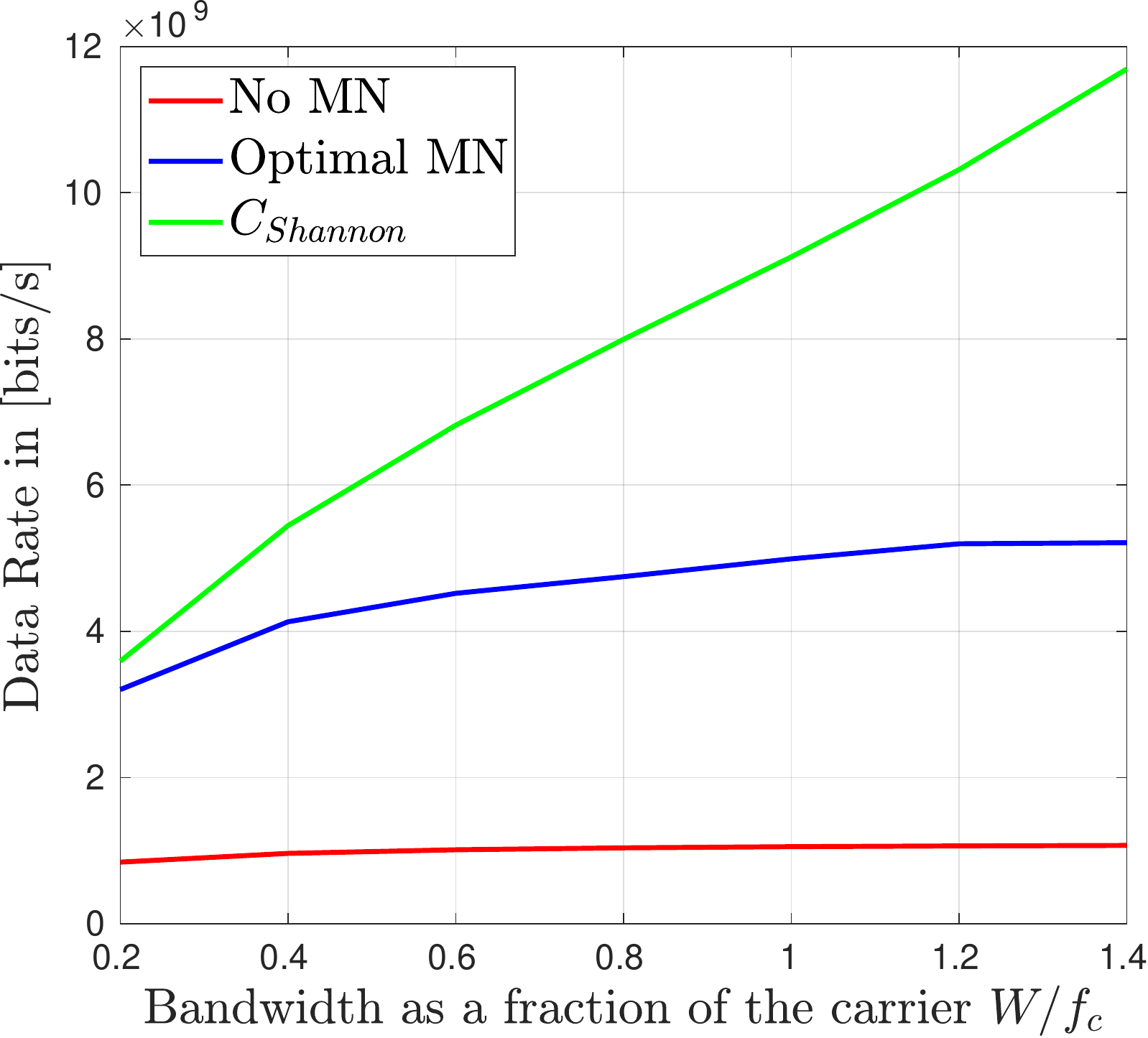}
  \caption{Middle-frequency regime: $f_c = 5\,[\textrm{GHz}]$,\,$P=\,4[W]$.}
  \label{fig:Capacity_bandwidth_2}
\end{subfigure}
\fboxsep=-\fboxrule\rule{0pt}{3cm}\hspace{4cm}
\begin{subfigure}{.5\textwidth}
  \centering
  \includegraphics[width=.65\linewidth]{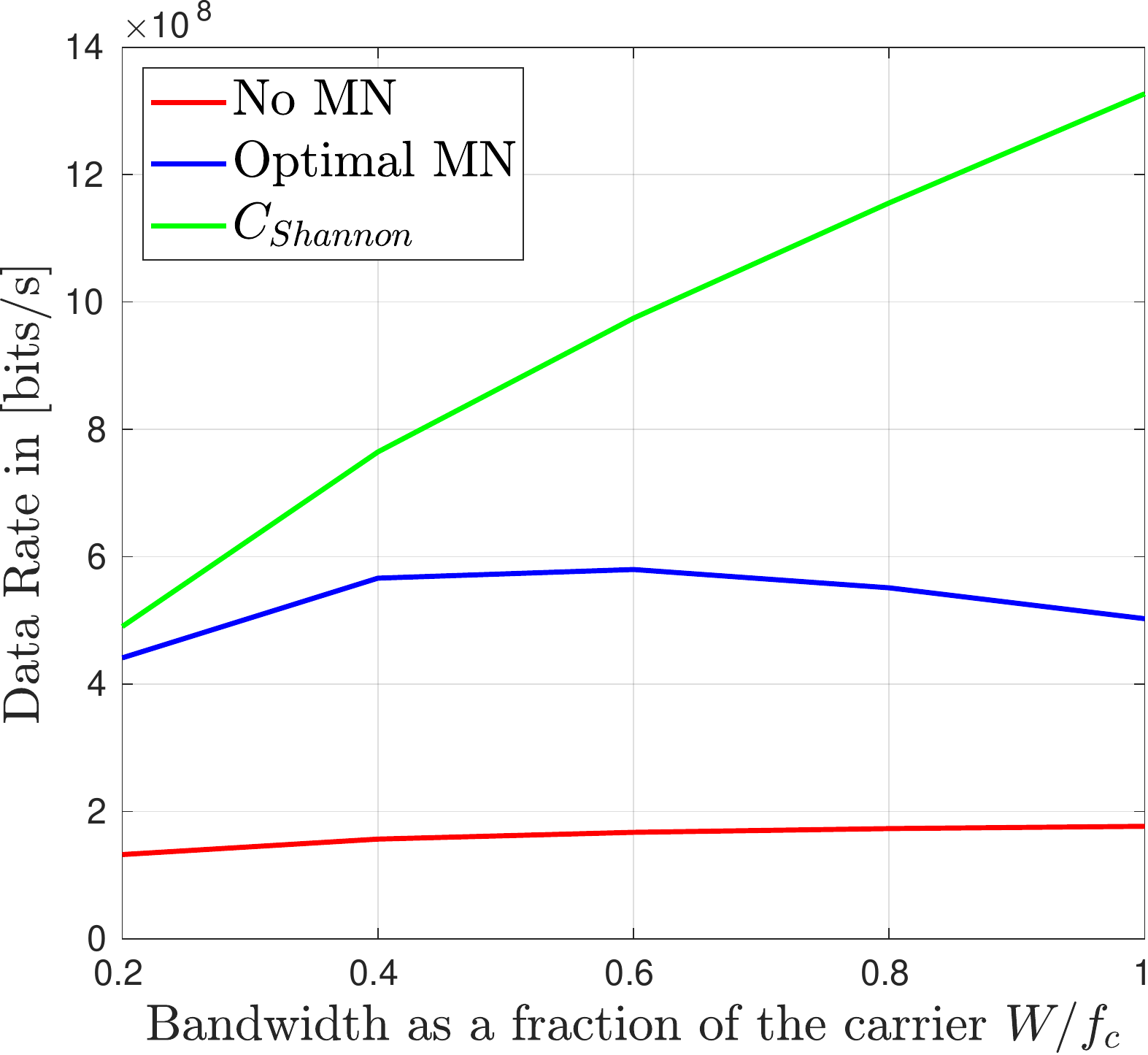}
  \caption{Low-frequency regime: $f_c = 600\,[\textrm{MHz}]$, $P=10\,[\textrm{mW}]$.}
  \label{fig:Capacity_bandwidth_3}
\end{subfigure}
\caption{Plots of the data rate as a function bandwidth measured as a fraction of the carrier $f_c$ for two different frequency regimes. The  antenna size is fixed to $\lambda/a = 20$.}
\label{fig:Signaling Bandwidth}
\end{figure}

\begin{figure}
    \centering
    \includegraphics[scale=0.4]{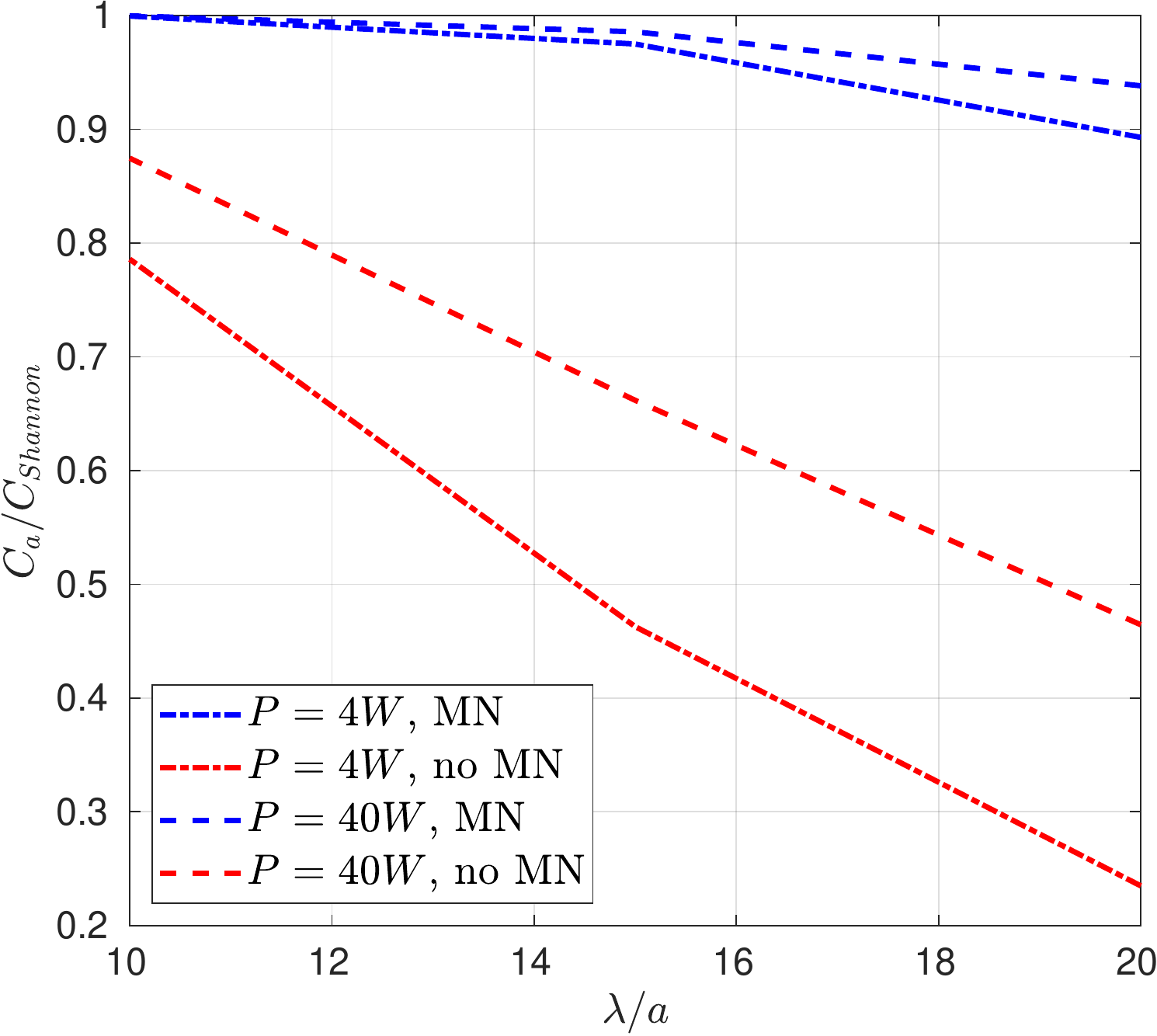}
    \caption{The fraction of the Shannon capacity at a fixed size $a$ as a function of the ratio $\lambda/a$ for two different values of transmit power. The carrier frequency is fixed to $f_c = 5\,[\textrm{GHz}]$ and the bandwidth to $0.2\,f_c$.}
    \label{fig:Trsnsmit_power} 
\end{figure}

We consider in Fig.~\ref{fig:antenna_size_capacity} the fraction of Shannon capacity that can be achievable for a given antenna size, $a$, as a function of the ratio $\lambda/a$. The mutual information $C_a$ is computed from (\ref{capacity}) with the optimized transmission coefficient from (\ref{optimal_transmission}) while the ideal mutual information is determined assuming frequency-flat antenna response, i.e.:
\begin{eqnarray}\label{Shannon}
C_{\rm{Shannon}} =  \int_{0}^ {\infty}\log_2\left(1 + \frac{P^*_t(f)\,|H(f)|^2}{N_0 + N_{\textrm{LNA}}}\right)\textrm{d}f~[\textrm{bits}/\textrm{s}].
\end{eqnarray}
For the same parameters of Fig.~\ref{fig:SNR_4_regimes}, we varied in Fig.~\ref{fig:antenna_size_capacity} the bandwidth, as a fraction of the carrier frequency $f_c$, from $0.2\,f_c$ to $0.8\,f_c$ where $f_c = 5\,[\textrm{GHz}]$. There, the antenna size decreases from left to right, from $a = 0.85\,[\textrm{cm}]$ at the left end where $\lambda/a = 7$ to  $a = 0.5\,[\textrm{cm}]$ at the right end where $\lambda/a = 12$. We observe the obvious fact that the achievable fraction of capacity decreases rapidly as the antenna size decreases. For example, for $\textrm{BW} = 0.4\,f_c$, we notice that the fraction of Shannon capacity that can be achieved drops by about $40\%$ as the antenna size decreases by a factor of 2 from $\lambda/a = 5$ to $\lambda/a = 10$. However, the optimal matching network can almost entirely remove the loss which is a significant improvement of multiple Gigabit in the data rate considering that the absolute $\textrm{BW} = 2[\textrm{GHz}]$. It also important to note that the normalized data rate achievable for a given $\lambda/a$ decreases by about $37\%$ with no matching at all as the bandwidth is increased $4$ times from $\textrm{BW} = 0.2\,f_c$ to $\textrm{BW} = 0.8\,f_c$. Such a decrease is expected since the broadband operation is known to be increasingly difficult when the antenna is compact in size w.r.t. the wavelength \cite{chu1948physical}.\\
We also notice the improvement brought by the use of the optimal MN even over a large bandwidth. It can be surprising to find out that the improvement stemming from the incorporation of a matching network is higher over larger bandwidth, e.g., about $66\%$ when $\textrm{BW} = 0.8\,f_c$ compared to $50\%$ when $\textrm{BW} = 0.2\,f_c$.\\ 
Fig. \ref{fig:Signaling Bandwidth} depicts the optimal signaling bandwidth and the effect of incorporating optimal matching network on the Shannon capacity given in (\ref{Shannon}), the maximum data rate achievable with the optimal matching network, as well as the one with no matching network as a function of bandwidth measured as a fraction of the carrier. There, the two sub-figures  are obtained in the low-frequency regime with $f_c = 600\,[\textrm{MHz}]$ as well as the medium-frequency regime with $f_c = 5\,[\textrm{GHz}]$. The apparent difference in the shape of the curves in Figs. (\ref{fig:Capacity_bandwidth_1}) and (\ref{fig:Capacity_bandwidth_2}) obtained in the low and medium-frequency regimes, respectively, can be attributed to the difference in the $\textrm{SNR}$ as the transmit power is kept fixed to $P=4\,[\textrm{W}]$, which means that a lower $\textrm{SNR}$ is experienced in the medium-frequency regime due to its larger absolute bandwidth.
We notice again that matching provides a substantial improvement both over smaller ($\textrm{BW}/f_c = 0.2$) and larger ($\textrm{BW}/f_c = 1$) bandwidths where the gain in rate stemming from utilizing a matching network is higher for larger bandwidths. However, the ideal performance, measured by the Shannon capacity, becomes more unrealistic as bandwidth increases according to Fano's theory \cite{Fano}. In fact, Fano made the observation that the matching tolerance decreases with bandwidth. From the range of values of the $\textrm{SNR}$ in Fig.~\ref{fig:SNR_4_regimes}, the data rate does not saturate, and hence an additional degradation of the $\textrm{SNR}$ could be further traded off for bandwidth. However, the best signaling with the optimum matching network can be seen in Fig.~\ref{fig:Capacity_bandwidth_3} by setting $P = 10\,[\textrm{mW}]$ and considering the lower $\textrm{SNR}$ regime. Fig.~\ref{fig:Capacity_bandwidth_2} also shows that matching has a particular utility in the low-$\textrm{SNR}$ regime, which is expected from the linear increase of the logarithm at low $\textrm{SNR}$. Finally, we explore the effect of $\textrm{SNR}$, or equivalently the transmit power $P$, on the performance of compact antennas. Fig. \ref{fig:Trsnsmit_power} shows the fraction of Shannon capacity that could be achieved for a given size $a$ as a function of $\lambda/a$.
There, two different power levels $P \in \{4,40\}\,[\textrm{W}]$ in the middle-frequency regime with $f_c = 5\,[\textrm{GHz}]$ and $\textrm{BW} = 0.2\,f_c$ are examined. It is seen that increasing the transmit power helps in achieving a larger fraction of the Shannon capacity even with no matching. For the smallest-size antenna, i.e., $\lambda/a = 20$, the improvement is almost three-fold. This again stems from the logarithmic dependence of the capacity at high $\textrm{SNR}$ where the SNR degradation due to the small antenna size does not play a big role, i.e., at most logarithmically.
\subsection{Interference scenario}
\begin{figure}
    \centering
    \includegraphics[scale=0.45]{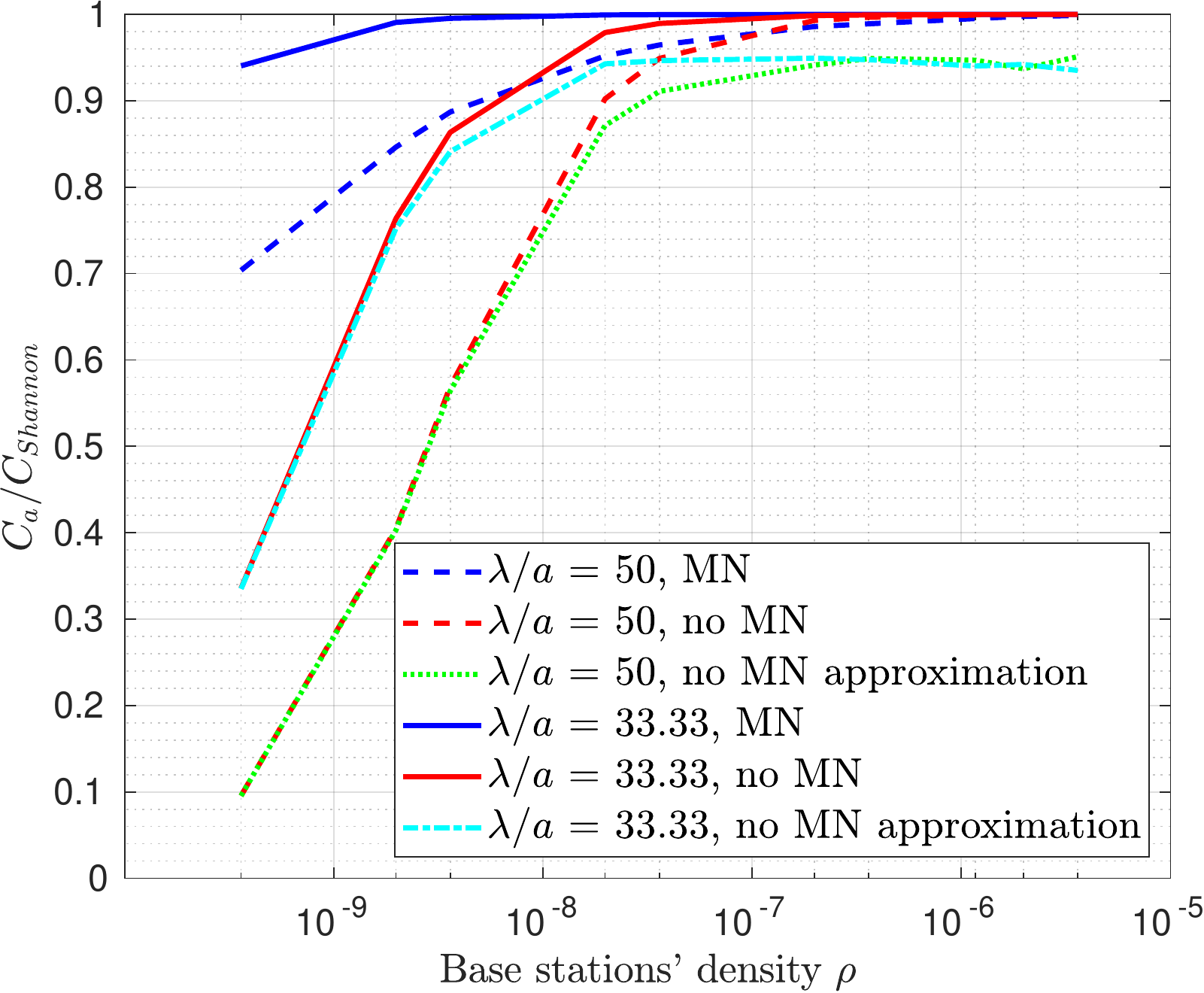}
    \caption{Achievable rate ratio vs. base stations' density $\rho$ with $\alpha=2.5$ for a fixed transmit power of $P=6\,[W]$.}
    \label{fig:Capacity_ratio_vs_user_density}
\end{figure}

We now investigate the impact of the antenna size and the optimal matching network on the data rate in presence of interference. The interference power is set to the same level as the transmit power which is now set to $P = 6~[W]$ and the pathloss exponent is set to $\alpha = 2.5$. We fix the distance between transmitter and receiver to be $d = R_0/3$, where $R_0$ represents an interference radius, such that there is only one user in a circular area of radius $R_0$. Additionally, the carrier frequency is fixed to $f_c = 600\,[\textrm{MHz}]$, the bandwidth to $\textrm{BW} = 0.25\,f_c$, and two small antenna sizes are considered, i.e., $\lambda/a \in \{50,33.33\}$.\\
Our approach is to compare the following two achievable rate ratios: $i)$ $C^{\textrm{no MN}}_{\textrm{inter}}/C_{\textrm{Shannon}}$ which represents the fraction of capacity achieved when no impedance matching is considered, and ii) $C^{\textrm{MN}}_{\textrm{inter}}/C_{\textrm{Shannon}}$ corresponding to the equivalent capacity fraction when the matching network is part of the SISO communication model.\\
Fig.~\ref{fig:Capacity_ratio_vs_user_density} depicts how these two ratios increase as a function of the user density $\rho$ for the aforementioned two antenna sizes. We observe that the proposed closed-form approximation (\ref{eq:closed-form-approx-expectation-log}) of $C^{\textrm{no MN}}_{\textrm{inter}}$ has a second-order truncation error that is rather small for this case with interference pathloss exponent $\alpha=2.5$ (but might be loose for higher $\alpha$). By varying the user density $\rho = 1/ (\pi \,R_0^2) \,[\textrm{users}/\textrm{m}^2]$ as a function of the cell radius $R_0$, it is seen that these achievable rate ratios approach one  with higher network densification, and reach the interference-limited regime (i.e., the achievable rate plateau) starting from a user density that depends on the presence or absence of the matching network.

In this interference-limited regime, it is interesting to notice that the achievable rate does not depend on the antenna size. This is because both the interference signal and the intended signal are undergoing the same antenna frequency response regardless of the antenna size.\\
When the interference power is large, the amplifier noise power can be safely ignored. In this case, the combined antenna/matching network frequency response cancels out from both the numerator and the denominator of the $\textrm{SNR}$, thereby rendering it identical to the Shannon capacity. It is useful to perceive the user density on the $x-$axis of Fig.~\ref{fig:Capacity_ratio_vs_user_density} by transforming it to the equivalent cell radius. Since there is only a single user in a cell of radius, $R_0 = 1/\sqrt{\rho\pi}$, the data point at $\rho = 10^{-8}\,[\textrm{users}/\textrm{m}^2]$ represents a cell of radius about $R_0\approx 5000 \,[\textrm{m}]$. The interference-limited regime occurs when $R_0\approx1000 \,[\textrm{m}]$, which is rather large, due to the low interference pathloss exponent.
Confirmed by Fig. \ref{fig:Capacity_ratio_vs_user_density} and the previous results, we conclude that future wireless networks open up new research directions for the antenna design, where information-theoretic design criteria might be more appropriate than conventional antenna design practices.

\section{Conclusion}
In this paper, we established the mutual information of a SISO wireless channel with the constraint on the antenna size at the receiver. After developing a circuit theoretic channel model, we computed the maximum mutual information per unit time between the input and output of a circuit system. Chu and Bode/Fano theories \cite{chu1948physical,Fano} were used to incorporate the size constraints. Our study revealed that the mutual information, specifically in the low-SNR regime, is severely degraded by the finite-size constraint. The optimal signaling bandwidth was further determined to be a significant fraction of the carrier showing that broadband communication systems are possible with compact antennas. Finally, we examined how the antenna size affects the mutual information in the noise-limited regime only, unlike the interference-limited regime where the antenna size is immaterial. Additional adaptations, tests, and experiments have been left for future work in which a generalization to multiple antennas and/or users can be considered. Future research should look at the practical realization of the derived matching network by means of the $N$th order reactive circuit. This is indeed similar in spirit to the original problem of coding. To achieve the derived limit in a practical system, the designs would require joint antenna/matching network optimization, which is another great direction for future investigation. To that end, the receive LNA model could be further refined to incorporate a more accurate noise model \cite{ivrlavc2010toward}. Finally, the analysis presented in this paper could be further extended to the case of MIMO systems. However, the design of the optimal matching network would be particularly challenging in the case of MIMO communication.
\begin{appendices}
\section{Derivation of Broadband matching constraints}\label{Fano_appendix}
In this appendix, based on the conditions of physical realizability of the reflection coefficient from \cite{Fano}, we derive the integral constraints in (\ref{constraint_1}) and (\ref{constraint_2}). It was shown by Darlington in \cite{darlington} that any physically realizable impedance can be regarded as an input impedance of a reactive two-port network that is terminated with a resistance $R$. For better illustration, the Darlington representation of the Chu-equivalent impedance is boxed in Fig.~\ref{fig:Fano_matching}. Without loss of generality, the resistance $R$ can be taken to be equal to $1\,\Omega$. To find the integral constraints, it is first required to determine the zeros of the transmission coefficient of Darlington network terminated with $1\,\Omega$ resistance as depicted in Fig.~\ref{fig:Darlington}. Using the Laplace transform, the transmission coefficient defined as:
\begin{equation}
   \widetilde{T}(s) = \frac{2\,V_1}{E_1},
\end{equation}
 can be found to be 
\begin{equation}\label{transmission_nomatch}
    \widetilde{T}(s) = \frac{2\,s^2\,\frac{a^2}{c^2}}{2\,s^2\,\frac{a^2}{c^2} + 2\,s\,\frac{a}{c} + 1}.
\end{equation}
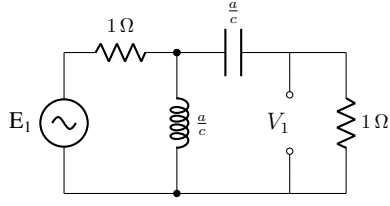
\begin{figure}[h!]
\centering
\begin{circuitikz}[american voltages, scale=0.75, every node/.style={transform shape}]
\draw (0,2.5) to[sV, l_=$\text{E}_{\text{1}}$, -] (0,0);
 \draw (0,0) -- (5,0);
 \draw (0,2.5) to[/tikz/circuitikz/bipoles/length=30pt, R, l=\mbox{\small{$1 \,\Omega$}}, -*] (2,2.5);
\draw (2,2.5) to[L, label=\mbox{\small{$\frac{a}{c}$}}, *-*] (2,0);
\draw (2,2.5) to[C, label=\mbox{\small{$\frac{a}{c}$}}, *-] (4,2.5);
\draw (4,2.5) -- (5,2.5);
\draw (4,2.5) to[short,-o] (4,1.75);
\draw (4,0) to[short,-o] (4,0.75);
 \draw (5,2.5)  to[/tikz/circuitikz/bipoles/length=30pt, R, l=\mbox{\small{$1 \,\Omega$}}, -] (5,0);
\node[] at (3.8,1.25) {$V_1$};
\end{circuitikz}
\caption{Darlington reactive network transmission coefficient}
\label{fig:Darlington}
\end{figure}

From \cite{Fano}, the two zeros of the transmission coefficient at the origin result in two integral constraints of the form:

\begin{subequations}
\small
    \begin{align}
        \frac{1}{2\pi^2} \int_{0}^{\infty} f^{-2} \ln\Bigg( \frac{1}{|\Gamma(f)|^2}\Bigg) \mathrm{d}f&=A_0^1-2 \gamma^{-1},\\
        \frac{1}{8\pi^4} \int_{0}^{\infty} f^{-4} \ln\Bigg( \frac{1}{|\Gamma(f)|^2} \Bigg)\mathrm{d}f&=-A_0^3+\frac{2}{3} \gamma^{-3}.
    \end{align}
\end{subequations}
where $A_0^1$ and $A_0^3$ have to be determined from the reflection coefficient of the Darlington-equivalent network terminated with $1\Omega$ resistance. This corresponds to finding the reflection coefficient from the input impedance in Fig.~\ref{fig:tm1} when setting $R_2 = 1$. The reflection coefficient is given by:
\begin{equation}\label{eq:no-matching-reflection-coeff}
    \widetilde{\Gamma}(s) = \frac{1}{2\,s^2\,\frac{a^2}{c^2} + 2\,s\,\frac{a}{c} + 1}.
\end{equation}
Moreover, the coefficients $A_0^1$ and $A_0^3$ are then given by:
\begin{subequations}
    \begin{align}
        A_0^1 &=  \frac{\textrm{d}}{\textrm{d}s}\ln\left(\frac{1}{\widetilde{\Gamma}(s)}\right)\bigg|_{s = 0} = \frac{2a}{c},\\
        A_0^3 &=  \frac{1}{6}\,\frac{\textrm{d}^3}{\textrm{d}s^3}\ln\left(\frac{1}{\widetilde{\Gamma}(s)}\right)\bigg|_{s = 0} = -\frac{4a^3}{3c^3},
    \end{align}
\end{subequations}
thereby yielding to the expressions in (\ref{constraint_1}) and (\ref{constraint_2}).
\section{Optimal reflection coefficient using the calculus of variation}\label{appendix:calculus-of-variation}
Using the definition of a local minimum of the Lagrangian $J(g)$ in (\ref{eq:augmented-lagragian}) at $g$, we have:
\begin{equation}
    \small
    J(g) \leq J(g + \epsilon\,\eta(f)) \triangleq \phi(\epsilon)
\end{equation}
where $\eta(f)$ is an arbitrarily shape function and $\epsilon$ represents the magnitude of variation. Setting the derivative of $\phi(\epsilon)$ to 0 around $\epsilon = 0$ yields:
\begin{equation}
\small
    \begin{aligned}
        0 = \frac{\textrm{d}\phi(\epsilon)}{\textrm{d}\epsilon}
        \bigg\rvert_{\epsilon=0}
        = \frac{\textrm{d}}{\textrm{d}\epsilon} J(g + \epsilon\,\eta(f)) \bigg\rvert_{\epsilon=0}&= \frac{\textrm{d}}{\textrm{d}(g + \epsilon\,\eta(f))}\,\frac{\textrm{d}(g + \epsilon\,\eta(f))}{\textrm{d}\epsilon} \, J(g + \epsilon\,\eta(f)) \bigg\rvert_{\epsilon=0}\\
        &= \frac{\textrm{d}\,J(g + \epsilon\,\eta(f))}{\textrm{d}(g + \epsilon\,\eta(f))} \, \eta(f) \bigg\rvert_{\epsilon=0}. 
    \end{aligned}
\end{equation}
Using the expression of $J(g)$ in (\ref{eq:augmented-lagragian}), we obtain
\begin{equation}\label{dphideps}
\small
    \begin{aligned}[b]
        \frac{\textrm{d}\phi(\epsilon)}{\textrm{d}\epsilon} \bigg\rvert_{\epsilon=0} &= \frac{\textrm{d}}{\textrm{d}\epsilon}\underbrace{\int_{0}^ {\infty}\,\frac{1}{\ln(2)}\,\ln\left(\frac{N_0(\mathcal{T}(f) + \epsilon\,\eta(f)) + N_{\textrm{LNA}} + P^*_t(f)\,|H(f)|^2\,(\mathcal{T}(f) + \epsilon\,\eta(f))}{N_0\,(\mathcal{T}(f) + \epsilon\,\eta(f)) + N_{\textrm{LNA}}}\right)\textrm{d}f}_{\triangleq\,\phi_1(\epsilon)}\\
        &\hspace{1cm}+ \frac{\textrm{d}}{\textrm{d}\epsilon}\Bigg[\underbrace{\mu_1\int_{0}^{\infty}f^{-2} \ln\bigg( \frac{1}{1 - \mathcal{T}(f) - \epsilon\,\eta(f)}\bigg)\, \mathrm{d}f \, -\mu_1\, K_1}_{\triangleq\,\phi_2(\epsilon)}\Bigg]\Bigg\rvert_{\epsilon=0}\\
        &\hspace{1cm}+ \frac{\textrm{d}}{\textrm{d}\epsilon}\Bigg[\underbrace{\mu_2\,\int_{0}^{\infty}  f^{-4} \ln\bigg( \frac{1}{1-\mathcal{T}(f)-\epsilon\,\eta(f)}\bigg)\, \mathrm{d}f - \mu_2\, K_2}_{\triangleq\,\phi_3(\epsilon)}\Bigg]\Bigg\rvert_{\epsilon=0}\\
        &\hspace{1cm}+ \mu_3\,\frac{\textrm{d}}{\textrm{d}\epsilon}\,\Big[\underbrace{\mathcal{T}(f) + \epsilon\,\eta(f)}_{\triangleq\,\phi_4(\epsilon)}\Big]\Big\rvert_{\epsilon=0} + \mu_4\,\frac{\textrm{d}}{\textrm{d}\epsilon}\,\Big[\underbrace{\mathcal{T}(f) + \epsilon\,\eta(f)}_{\triangleq\,\phi_5(\epsilon)}\Big]\Big\rvert_{\epsilon=0}.
    \end{aligned}
\end{equation}
After computing the derivatives of $\phi_i(\epsilon)$ for $i \in \{1,2,3,4,5\}$ and letting $\epsilon = 0$, we get:

\begin{equation}
\small
    \begin{aligned}[b]
         \frac{\textrm{d}\phi(\epsilon)}{\textrm{d}\epsilon} \bigg\rvert_{\epsilon=0} &= \frac{\textrm{d}\phi_1(\epsilon)}{\textrm{d}\epsilon} \bigg\rvert_{\epsilon=0} +  \frac{\textrm{d}\phi_2(\epsilon)}{\textrm{d}\epsilon} \bigg\rvert_{\epsilon=0} +  \frac{\textrm{d}\phi_3(\epsilon)}{\textrm{d}\epsilon} \bigg\rvert_{\epsilon=0} +  \frac{\textrm{d}\phi_4(\epsilon)}{\textrm{d}\epsilon} \bigg\rvert_{\epsilon=0} + 
         \frac{\textrm{d}\phi_5(\epsilon)}{\textrm{d}\epsilon} \bigg\rvert_{\epsilon=0}\\
         &=\int_{0}^{\infty}\eta(f)\, \underbrace{\frac{1}{\ln(2)}\frac{N_{\textrm{LNA}}\, P^*_t(f)\,|H(f)|^2}{\big((N_0 + P^*_t(f)\,|H(f)|^2)\,\mathcal{T}(f) + N_{\textrm{LNA}}\big) \big( N_0\,\mathcal{T}(f) + N_{\textrm{LNA}}\big)}}_{\triangleq\,\psi_1(f)}\,\textrm{d}f\\
         &\hspace{0.5cm}+ \Bigg(\int_{0}^{\infty} \eta(f)\,\underbrace{\mu_1\,\frac{f^{-2}}{1-\mathcal{T}(f)}}_{\triangleq\,\psi_2(f)}\mathrm{d}f - \mu_1 \frac{\textrm{d}}{\textrm{d}\epsilon}\,K_1\Bigg) + \Bigg(\int_{0}^{\infty} \eta(f)\,\underbrace{\mu_2\,\frac{f^{-4}}{1-\mathcal{T}(f)}}_{\triangleq\,\psi_3(f)}
         \mathrm{d}f - \mu_2\,\frac{\textrm{d}}{\textrm{d}\epsilon}\,K_2\Bigg)\\
         &\hspace{0.5cm}+ \underbrace{\mu_3}_{\triangleq\,\psi_4(f)}\,\eta(f) + \underbrace{\mu_4}_{\triangleq\,\psi_5(f)}\,\eta(f).
    \end{aligned}
\end{equation}
Assuming $K_1$ and $K_2$ are constant and therefore independent of $\epsilon$, it follows that
\begin{equation}
    \small
    \int_{0}^ {\infty} \eta(f)\big(\psi_1(f) + \psi_2(f) + \psi_3(f) + \psi_4(f) + \psi_5(f)\big) \mathrm{d}f = 0
\end{equation}
Since the above integral should vanish for any arbitrary function $\eta(f)$, we must have
\begin{equation}\label{eq:condition-sum-equal-0}
  \small
  \psi_1(f) + \psi_2(f) + \psi_3(f) + \psi_4(f) + \psi_5(f) = 0  
\end{equation}
To obtain a second-order polynomial, we set $\mu_3=\mu_4=0$. After substituting the expressions of $\psi_i(f)$ for $i \in \{1,2,3,4,5\}$ in (\ref{eq:condition-sum-equal-0}) and setting the numerator to 0, we get a quadratic polynomial whose coefficients are given in (\ref{quadratic solution}).

\section{Derivation of $\mathcal{T}^{\tiny{\starletfill}}(f)$ in (\ref{optimal_transmission})}\label{appendix3:derivation-optimal-transmission-coeff}
Consider the constrained maximization in \ref{capacity} and \ref{signum_function}.
We first relax the equality constraints (\ref{signum_function}) to inequality constraints as follow:
\begin{subequations}
\label{appendix3:signum_function_inequality}
    \small
    \begin{align}
        &\int_{0}^{\infty} f^{-2} \ln\left(\frac{1}{1 - \mathcal{T}(f)}\right) \mathrm{d}f\leq K_1,\\
        & \int_{0}^{\infty} f^{-4} \ln\left( \frac{1}{1-\mathcal{T}(f)} \right)\mathrm{d}f \leq K_2.
    \end{align}
\end{subequations}
which does not impact the  optimization result, since the inequalities hold with equality at the optimum (because larger antenna always improves performance). 
The Lagrangian associated with (\ref{capacity}) and (\ref{appendix3:signum_function_inequality}) is given by:
\begin{equation}\label{appendix3:Lagrangian}
    \small
    \begin{aligned}[b]
        J(\mathcal{T}(f), \gamma, \mu_1, \mu_2,  \mu_3(f), \mu_4(f)) &= \int_{0}^ {\infty}\log_2\left(1 + \frac{P^*_t(f)\,|H(f)|^2\,\mathcal{T}(f)}{N_0\,\mathcal{T}(f) + N_{LNA}}\right)df \\
        &\hspace{1cm}+ \mu_1\left(\int_{0}^{\infty} f^{-2} \ln\bigg( \frac{1}{1 - \mathcal{T}(f)}\bigg) \mathrm{d}f - K_1\right) \\
        &\hspace{2cm}+ \mu_2\left(\int_{0}^{\infty} f^{-4} \ln\bigg( \frac{1}{1-\mathcal{T}(f)}\bigg) \mathrm{d}f - K_2\right) \\
        &\hspace{3cm}+ \mu_3(f)\,\mathcal{T}(f) + \mu_4(f) (\mathcal{T}(f)-1),
    \end{aligned}
\end{equation}
with $\mu_1\leq 0, \mu_2\leq 0, \mu_4(f)\leq 0~\forall f$, and $\mu_3(f)\geq 0~\forall f$. For $\mu_3(f)=\mu_4(f)=0$ (inactive constraints $0 \leq \mathcal{T}^{\tiny{\starletfill}}(f) \leq 1$), using variational calculus-based optimization, it is shown in Appendix \ref{appendix:calculus-of-variation} that the gradient condition of (\ref{appendix3:Lagrangian}) yields:
\begin{equation}\label{appendix:condition-mu3=mu4=0}
    \small
    C_1(f)\mathcal{T}^{\tiny{\starletfill}}(f)^2+ C_2(f)\mathcal{T}^{\tiny{\starletfill}}(f)+C_3(f)=0 ~ \textrm { for } \mu_3(f)=\mu_4(f)=0
\end{equation}
where 
\begin{subequations}\label{appendix3:quadratic solution}
    \small
    \begin{align}
    &\hspace{-0.25cm}C_1(f) = (N_0 + P^*_t(f)\,|H(f)|^2)\,N_0\,(\mu_1\,f^{-2} + \mu_2\,f^{-4}), \\
    &\hspace{-0.25cm}C_2(f) = (2\,N_0\,N_{\textrm{LNA}} + N_{\textrm{LNA}}\,P^*_t(f)|H(f)|^2)\,(\mu_1\,f^{-2} + \mu_2\,f^{-4}) - P^*_t(f)\,|H(f)|^2\,N_{\textrm{LNA}}, \\
    &\hspace{-0.25cm}C_3(f) = P^*_t(f)\,|H(f)|^2\,N_{\textrm{LNA}} + N_{\textrm{LNA}}^2\,(\mu_1\,f^{-2} + \mu_2\,f^{-4}).
    \end{align}
\end{subequations}
Setting the derivative of the Lagrangian (\ref{appendix3:Lagrangian}) w.r.t. $\gamma$ to zero gives:
\begin{equation}\label{appendix:optimal-gamma}
\small
\begin{aligned}
    0 = - \mu_1\,\frac{\textrm{d}}{\textrm{d}\gamma} K_1 - \mu_2\,\frac{\textrm{d}}{\textrm{d}\gamma} K_2 = \frac{\mu_1}{\gamma^2}  - \frac{4\pi^2}{\gamma^4}.
\end{aligned}
\end{equation}
By recalling that $\gamma$ is the positive real-valued zero of the reflection coefficient, i.e. $\gamma >0$, we find that its optimal value is $\gamma = 2\pi\sqrt{\frac{\mu_2}{\mu_1}}$.
Finally, using (\ref{appendix:condition-mu3=mu4=0}) and (\ref{appendix:optimal-gamma}), we obtain the following nine KKT conditions of the Lagrangian in (\ref{appendix3:Lagrangian}):
\begin{equation*}\small
    \begin{aligned}
        &\circled{1} ~C_1(f)\mathcal{T}^{\tiny{\starletfill}}(f)^2+ C_2(f)\mathcal{T}^{\tiny{\starletfill}}(f)+C_3(f)=0~\textrm { for } \mu_3(f)=\mu_4(f)=0\\
        &\circled{2} ~\mu_1\left(\int_{0}^{\infty} f^{-2} \ln\bigg( \frac{1}{1 - \mathcal{T}(f)}\bigg) \mathrm{d}f - K_1\right) =0, \quad \circled{3}~\mu_2\left(\int_{0}^{\infty} f^{-4} \ln\bigg( \frac{1}{1-\mathcal{T}(f)}\bigg) \mathrm{d}f - K_2\right)=0 \\   
        &\circled{4}~\mu_3(f)\,\mathcal{T}(f)^{\tiny{\starletfill}}=0~ \forall f ,\quad \circled{5} ~\mu_4(f)\,(\mathcal{T}(f)^{\tiny{\starletfill}}-1)=0 ~\forall f,\quad \circled{5}~\mu_1\leq 0, ~\mu_2\leq 0,~ \mu_4(f)\leq 0~\forall f\\
        &\circled{7}~\mu_3(f)\geq 0~\forall f,\quad \circled{8}~0 \leq\mathcal{T}(f)^{\tiny{\starletfill}} \leq 1~\forall f, \quad \circled{9} ~\gamma = 2\pi\sqrt{\frac{\mu_2}{\mu_1}}
    \end{aligned}
\end{equation*}
Since both Fano inequalities hold with equality, we have $\mu_1<  0$ and $\mu_2<  0$.
We therefore conclude that $C_1(f) < 0$,  $C_2(f) < 0$ as well as $C_1(f)+C_2(f)+ C_3(f) < 0  $. Consequently, the first KKT condtion cannot have a solution $\mathcal{T}^{\tiny{\starletfill}}(f) \geq 1$ ({\em Proof:} Consider the function $f(x)=C_1(f)x^2+C_2(f)x+C_3(f)$. We have $f(1)<0$ and $f'(x) \leq 0, ~\forall~x >0 $, thus the continuous function $f(x)$ does not have a root over the interval $x \geq 1$). Therefore, the constraint $\mathcal{T}(f) \leq 1$ is always inactive at the optimum (i.e. $\mathcal{T}^{\tiny{\starletfill}}(f) < 1$ ) and $\mu_4(f)=0 ~\forall f$.

Now, solving for $\mathcal{T}^{\tiny{\starletfill}}(f)$ the optimal reflection coefficient is given by the solution of the quadratic equation, after enforcing the condition that $\mathcal{T}^{\tiny{\starletfill}}(f) \geq  0$, i.e.:

\begin{equation*}
\label{appendix3:optimal_transmission}
\small
    \mathcal{T}^{\tiny{\starletfill}}(f) ~=~ \max \Bigg(0,\frac{-C_2(f) \pm  \sqrt{C_2^2(f) - 4\,C_1(f)\,C_3(f)}}{2\,C_1(f)}\Bigg).
\end{equation*}
Finally, we can ignore the solutions with the plus sign as it would lead always to negative values, since $C_1(f) < 0$,  $C_2(f) < 0$, thereby yielding:
\begin{equation*}
\label{appendix3:optimal_transmission2}
\small
    \mathcal{T}^{\tiny{\starletfill}}(f) ~=~ \max \Bigg(0,\frac{-C_2(f) -   \sqrt{C_2^2(f) - 4\,C_1(f)\,C_3(f)}}{2\,C_1(f)}\Bigg).
\end{equation*}

\end{appendices}
\bibliographystyle{IEEEtran}
\bibliography{IEEEabrv,references}
 
\end{document}